# PADÉ FILTERING,
# Principles and Use:
# an Introductory Report

Jean-Daniel Fournier[*][†]
and
Mikhaël Pichot du Mézeray[♦][‡]

Cannes and Nice, Fall 2024

Short abstract

This report aims to provide gravitational waves data analysts with an introduction to the ideas and practice of the Padé Filtering method for disentangling a signal from the noise. Technically it comes to the tracking of the zeros and singularities of random z-Transforms by noisy Padé Approximants.

Extended abstract see next page.

Acknowledgments

P. Astone, P. Barone, B. Beckermann, D. Bessis, M.-A. Bizouard, N. Christiensen, Th. Cokelaer, E.Cuoco, I. Fiori, L. Perotti, E. Slezak, G. Turchetti, D. Verkindt, J.-Y. Vinet.

And in memory of J. Gilewicz, J.-P. Kahane, M. Pindor.

---

[*] Retired CNRS scientist, « collaborateur bénévole » of the « Observatoire de la Côte d'Azur »
[†] Member of the Virgo collaboration
[♦] ARTEMIS laboratory, CNRS and OCA, UCA
[‡] Member of the Virgo collaboration

# EXTENDED ABSTRACT

This report aims to provide gravitational waves data analysts with an introduction to the ideas and practice of the Padé Filtering (PF) method for disentangling a signal from the noise. Technically it comes to the tracking of the zeros and singularities of random z-Transforms by noisy Padé Approximants (PA); the associated zeros and poles separate into classes in $\mathbb{C}$, reflecting the signal and the noises, hence the Padé filtering tools. In this document the motion, separation, pairing (Froissart doublets) and condensation of roots of certain random polynomials occurring in data analysis are first reviewed. Then, mainly new, PF studies are exposed concerning (i) prototypical signals and noises; (ii) signals and noises occurring in the interferometric detection of gravitational waves. Details on actual implementation and a proof of concept using Virgo data are provided as well as a reminder appendix on classical PA's.

# PADÉ  FILTERING
## Principles and Use:
## An Introductory Report

# PART A

# INTRODUCTION

I.      Main themes and outline
1.   Main themes
2.   Outline

II.     Going into the complex
3.   Motives in the real versus patterns of poles in the complex
4.   Two odd probabilistic phenomena in the complex
a. The Kac phenomenon
b. The Froissart phenomenon

III.    Adventures of physics' complex tools in probability land
5.  The random z-Transform
a.   Gaussian white case
b.   The random autocorrelation and the Wiener-Khintchine theorem

# PART A: INTRODUCTION

## I. Main themes and outline

1. Main themes

Suppose we are given a set $\mathcal{D}_N$ of $N$ real data, deterministic or noisy. We consider its generating function, called z-Transform in the Data Analysis (DA) framework; we use the following notations,

$$\mathcal{D}_N = \{ d_0, d_1, \ldots, d_{N-1} \},$$

$$T_N^{(\mathcal{D})}(z) = \sum_{m=0}^{N-1} d_m \, z^m.$$

We observe that this Transform is nothing but the analytic continuation of the Discrete Fourier Transform (DFT) of $\mathcal{D}_N$,

$$F_N^{(\mathcal{D})}(\theta) = \sum_{m=0}^{N-1} d_m \, e^{im\theta} \ ,$$

in the complex variable

$$z = e^{i\theta}.$$

The normalized modulus squared of the DFT,

$$\mathcal{E}_N^{(\mathcal{D})}(\theta) = \frac{1}{N} \left| F_N^{(\mathcal{D})}(\theta) \right|^2, \ \ \theta \in \mathbb{R} \, ,$$

is called the Fourier Energy Spectrum and is a favorite tool in DA; one usually characterizes its profile in terms of its maxima, called peaks or lines, and of its behavior for small $|\theta|$. We note that it can be extended to $z$ not belonging to the unit circle $\mathcal{C}(0,1)$ via the formula

$$\mathcal{E}_N^{(\mathcal{D})}(z) = \frac{1}{N} \, T_N^{(\mathcal{D})}(z) \ T_N^{(\mathcal{D})} \left( \frac{1}{z} \right).$$

In the present work we advocate that

(i) *it is much more advisable and reliable to characterize $\mathcal{D}_N$ using the complex analytic structure of its z-Transform rather than the real peaks of its Fourier Energy Spectrum;*

(ii) *this holds both in the deterministic and the random case, with appropriate definitions and provisos.*

We demonstrate that

(iii) *the rational (possibly random) Padé Approximants (PA's) to the z-Transform conveniently provide (the statistics of) patterns of complex zeros and poles pointing to the true analytic structure of T(z) and in turn to the major (physical) features of the data.*

As to the practical use of these principles,

(iv) *we focus on academic data and on gravitational waves (GW) data, collected or expected at the Virgo, LIGO, KAGRA and LISA detectors.*

## 2. Outline

Our main motivation in composing this introductory report is to arouse the interest of the DA practitioners, especially those involved in GWDA. For this reason, we have divided the material into three clearly different parts[§].

In Part A, chapters II and III, we try to build up the intuition of the reader concerning complex and/or probabilistic phenomena using elementary or classical mathematical notions and facts.

In Part B we give an update on the distribution of the complex zeros/poles of real polynomials and rational approximants or transfer functions in a probabilistic setting. This includes the frequency detection via the motion of the complex roots of random orthogonal Szegö polynomials of degree $n$ ; those polynomials are derived from the data set $\mathcal{D}_N$ and the study is made in a large $n$ and $N$ asymptotics. Also we discuss the relationship between the $z$-Transform and the Fourier Energy Spectrum of, discrete time, white and colored, gaussian processes [**] ; we perform thorough calculations in the white case and pink and red cases[††]. Some results are ours, hitherto not published. Some are by JDF with other coworkers or by other groups and some date back to the years 1990's or 2000's. In the perspective of contemporary Padé Filtering research, they retrospectively appear as pioneering works but they are not logical prerequisites for this report's main topic. In chapter V, we examine the analytic structure of Padé Approximants and other Rational Approximants to noisy generating functions. The Froissart doublets phenomenon described in chapter II as a curiosity is generalized; it appears to be the key to the separation between patterns of zeros/poles reflecting the presence of a deterministic signal and patterns reflecting the presence of randomness; white and colored noises are also distinguished by distinctive patterns.

The Padé Filtering (PF) method can be summarized by "use PA's to look for the complex singularities of the z-Transform of your data". In part C it is addressed from a historical, scientific and practical perspective. We emphasize the information compression operated by PF, as perceived by D. Bessis in his JCAM (1996) paper. We prove the validity of the concept on actual Virgo data. We use the concept in the analysis of signals and noises at GW interferometric antennas. In the final (§17) paragraph we go further on implementing and questioning the method.

The document closes with four complements and annexes, including a reminder on classical Padé Approximants theory, and one on basic probability calculations. We assume that the reader of the main body of the document is familiar with their content. We have also a bibliography with some comments and the captions of the figures.

All in all, this purposely pedagogical report displays features of a review paper as well as of a research paper. We hope this mix will be useful to people from mathematical physics and pure and applied mathematics as well as to the DA specialists.

---

[§] Observe however that here and there we have inserted remarks or notes pointing to open mathematical issues; for instance, in Part C, concerning the local behavior of probabilistic measures of pole's positions in the complex.
[**] ARMA generated processes, see Section 8 for definitions.
[††] Respectively AR(1) and AR(2) generated. See section 8.

## II.    Going into the complex

### 3.    Motives in the real versus patterns of poles in the complex.

Using the partial fraction decomposition on $\mathbb{C}$, a set of poles together with their amplitudes can be associated to any rational function. On some given rational physical laws - or rational approximants thereof -, one can explore the relationship between this complex description and the description in the real.

Here we consider first the simple real function

$$L_I(w) = \frac{1}{\pi} \frac{\Gamma}{(w-w_0)^2 + \Gamma^2} \;,$$

where $w$ is a real variable, while $\Gamma$ and $w_0$ are positive parameters. In atomic physics [‡‡], $w$ is a frequency and $L_I(w)$ is known as the Lorentz shape; it describes a spectrum with one single line centered at $w_0$ and natural or collisional half width $\Gamma$. This function is normalized in the sense that its integral is one; its maximum $\frac{1}{\pi} \frac{1}{\Gamma}$ is reached at $w_0$. This function being real-analytic, there is a single pair of two complex conjugate simple poles located at $w_\pm = w_0 \pm i\Gamma$, with residues $R_\pm = \mp \frac{i}{2\pi}$. The shape of the graph for various $\Gamma$ is depicted in Fig. 3.1.

We consider now a two lines spectrum $L_{II}(w)$, which we model as the half sum of two Lorentz profiles with their respective maxima located at $w_1$ and $w_2$ and the same half width $\Gamma$. This spectrum thus reads

$$L_{II}(w) = \frac{\Gamma}{2\pi} \left[ \frac{1}{(w-w_1)^2 + \Gamma^2} + \frac{1}{(w-w_2)^2 + \Gamma^2} \right].$$

The positions of the extrema of its shape are the real solutions of the condition

$$\frac{dL_{II}(w)}{dw} = 0 \;.$$

We introduce a notation for the distance - in frequency domain – between the two original lines,

$$w_2 - w_1 \equiv S\,\Gamma,$$

where the original half width $\Gamma$ is taken as unit frequency. The solutions of (3.3) are first

$$w_e = \frac{1}{2}(w_1 + w_2)$$

for any S $\geq$ 0. In addition, provided

$$S \geq \frac{2}{\sqrt{3}} \approx 1.1547 \;,$$

there is a couple of extrema $(w_{e-}, w_{e+})$ located at

$$w_{e+} = w_1 + \frac{1}{2}\Gamma\,S\,\mathfrak{f}(S)$$
$$w_{e-} = w_2 - \frac{1}{2}\Gamma\,S\,\mathfrak{f}(S)$$
$$\mathfrak{f}(S) = 1 + (1 + \frac{4}{S^2})^{\frac{1}{4}} \left[ 2 - (1 + \frac{4}{S^2})^{\frac{1}{2}} \right]^{\frac{1}{2}}$$

---

with the configuration

$$w_{e-} \leq w_e \leq w_{e+} .$$  (3.8)

For $S = 2/\sqrt{3}$ , the three coalesce. For $0 < S < 2/\sqrt{3}$ , the couple $(w_{e-}, w_{e+})$ disappears and one is left with only one extremum, $w_e$.

The scenario, governed by the parameter $S$, is thus as follows. For sufficiently separated original lines, the new spectrum exhibits two maxima, reminiscent of the two lines; between those two maxima, there is a minimum at $w_e$ . For closer lines, the couple $(w_{e-}, w_{e+})$ disappears; the extremum $w_e$ becomes the unique global maximum. Fig. 3.2 illustrates the different phases of this scenario.

At the same time, the topology of the complex analytic structure of $L_{II}(w)$ is stable: there are always two pairs of complex conjugate simple poles. Quantitatively the poles are located at $w_{1,2} \pm i\Gamma$ and do have invariant residues $\frac{1}{2}\frac{1}{2\pi}(\mp i)$. For fixed $w_{1,2}$, they just go deeper in the complex when the half width $\Gamma$ increases.

Figure 3.1                    Figure 3.2

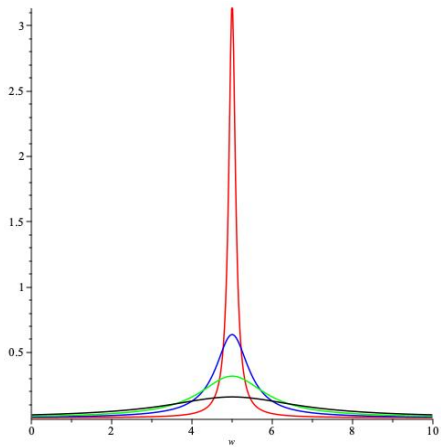    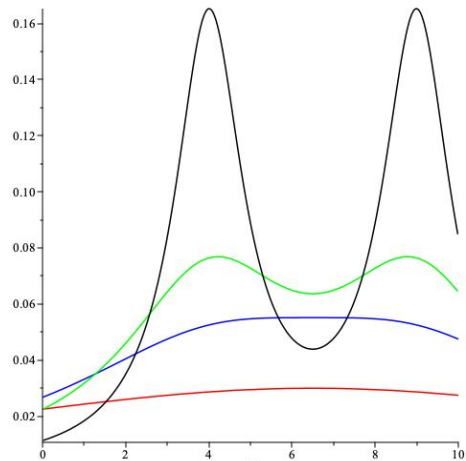

$w_0 = 5$                      $w_1 = 4$        $w_2 = 9$
$\Gamma = 0.1$    red          $\Gamma = 1$        black
$\Gamma = 0.5$    blue         $\Gamma = 2.5$      green
$\Gamma = 1$      green        $\Gamma = 4.3301$   blue
$\Gamma = 2$      black        $\Gamma = 10$       red

*The moral of this elementary exercise is that the coding of the physics behind a profile via the complex analytic structure is more faithful than the coding via the real peaks.*

4.  Two odd probabilistic phenomena in the complex

In this paragraph we assume that the reader is somehow familiar with probability theory and deterministic Padé approximation.

a.  The Kac phenomenon. We consider the real roots of a real random polynomial

$$P_n(z) = \sum_{k=0}^{n} a_k \, z^k \,,$$

where the coefficients $a_k$ are real random gaussian variables. One observes that, for given real $z$, $P_n(z)$ is itself a real gaussian variable and thus that, via the manipulation of gaussian integrals, the average density $\rho_n(t)$ along $\mathbb{R}$ of its real roots is amenable to actual calculation. Let us here admit the result in the case where the random real $a_k$ are independent and identically distributed according to the gaussian law with zero mean and variance unity. The average density along the real axis then reads

$$\rho_n(t) = \frac{1}{\pi} \left[ \frac{1}{(1-t^2)^2} - \frac{(n+1)^2 \, t^{2n}}{(1-t^{2n+2})^2} \right]^{\frac{1}{2}} \,.$$

For any fixed $n$, this real positive quantity has a positive finite integral over $\mathbb{R}$, which is the average number $\mathbb{E}[N_n]$ of real roots of $P_n(t)$; for large $n$, this number behaves according to

$$\mathbb{E}[N_n] \underset{n \gg 1}{\simeq} \frac{2}{\pi}\ln n \,.$$

See Fig.4.3. Thus for any fixed $n$, a sizeable part of the roots of $P_n$ lies on the real axis, although the latter has a zero Lebesgue measure in the plane; this was first pointed out by M. Kac (1943), see fig (4.3). It is also noteworthy that $\rho_n(t)$ is peaked very near $\pm 1$, a phenomenon which is consistent with the peak on the unit circle of the density of complex roots of real random homogeneous polynomials, see §7. See figures (4. 1) and (4.2); see also proofs and details in the lecture by B. Dujardin at the Porquerolles School (2003). We finally observe that randomness creates order: little can be said on the real roots of a general deterministic polynomial, while the random case exhibits the Kac phenomenon and is amenable to calculation.

Figure 4.1                                  Figure 4.2

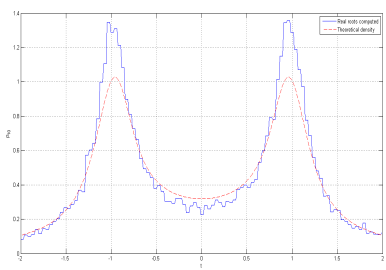

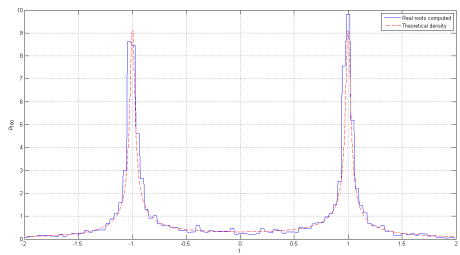

Figure 4.3

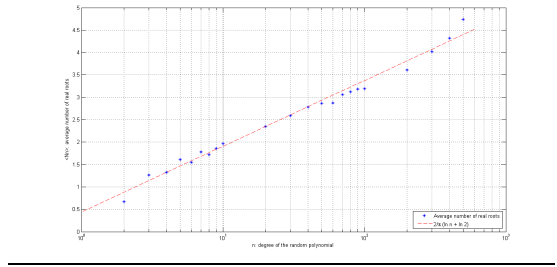

b.    The Froissart phenomenon. See chapter IX and chapter V for precise definitions and properties. In the deterministic setting, a Padé Approximant is a rational function

$$PA\,(z) = \frac{P(z)}{Q(z)}$$

which is an approximation of a function known by its Taylor series

$$f(z) = a_0 + a_1 z + \cdots + a_n z^n + \cdots$$

The approximation condition is written as the matching of the Taylor series of $f(z)$ and $PA\,(z)$ up to a certain order. To look at the effect of randomness, let us consider the noisy

$$f_\varepsilon(z) = a_0(1 + \varepsilon r_0) + a_1(1 + \varepsilon r_1)z + \cdots + a_n(1 + \varepsilon r_n)z^n + \cdots$$

and its Padé Approximants

$$PA_\varepsilon(z) = \frac{P_\varepsilon(z)}{Q_\varepsilon(z)};$$

$\varepsilon$ is a governing small parameter and the $r_n$ are random real numbers. For many classes of series, and various assumptions on the noise, the following takes place concerning the roots of $P_\varepsilon(z)$ and $Q_\varepsilon(z)$ in a given realization:

(i)     some roots of $Q_\varepsilon$ reflect the singularities of $f(z)$ while others are at a $O(\varepsilon)$ distance from the unit circle in $\mathbb{C}$

(ii)    some roots of $P_\varepsilon$ reflect the zeros of $f(z)$ while others are at a $O(\varepsilon)$ distance from the unit circle in $\mathbb{C}$

(iii)   the poles and zeros of $PA_\varepsilon(z)$ having a modulus close to 1 can be grouped into zero-pole pairs or doublets named after M. Froissart who first noticed them (1969).

These doublets, located near the unit circle, have an $O(\varepsilon)$ extension. They appear in various noisy rational approximants and are sort of a footprint of the noise (chapter V and chapter VI).

## III.    Adventures of physics' complex tools in probability land

### 5.    The random z-transform

Il there is some randomness in the data, so is it too in the z-transform (see §1). One is then bound to ask probabilistic questions, concerning expectation values, moments, … of the random $T_N^{(\mathcal{D})}(z)$. In this paragraph we explore some of them.

We suppose the data are real random variables (rrv) independent and identically distributed (iid) according to the normal law $\mathcal{N}(\mu, \sigma)$ with mean $\mu$ and variance $\sigma^2$ ; that is the probability distribution function $pdf\,(d_m)$ of the $m-th$ variable $d_m$ is

$$pdf\,(d_m) = \frac{1}{\sqrt{2\pi}}\frac{1}{\sigma}\,exp\left[-\frac{1}{2}\left(\frac{d_m - \mu}{\sigma}\right)^2\right],$$

$$\mu \in \mathbb{R}, \sigma \in \mathbb{R}_+^* \text{, both independent of } m,$$

and there is no correlation between $d_m$ and $d_n$ if $m \neq n$. We introduce reduced variables $\{\delta_0, \ldots, \delta_m, \ldots, \delta_{N-1}\}$ such that

$$d_m = \mu + \sigma\delta_m .$$

The $\delta_m$'s are rrv, iid according to $\mathcal{N}(0,1)$ and, for any $N$, the z-transform of the $d_m'$ can be rewritten

$$T_N^{\{d\}}(z) = \mu + z\mu + \cdots + z^{N-1}\mu + \sigma[\delta_0 + \delta_1 z + \cdots + \delta_{N-1}z^{N-1}]$$

$$= \mu\frac{1-z^N}{1-z} + \sigma T_N^{\{\delta\}}(z).$$

The probabilistic problem is thus reduced to the study of the random z-transform $T_N^{\{\delta\}}(z)$ of the set $\{\delta_m\}$. In the sequel we will stick with the notation $d_m$, assuming

$$\mathbb{E}[d_m]=0,$$

$$\mathbb{E}[d_m{}^2] = 1,$$

$$\mathbb{E}[d_m d_n] = 0, m \neq n ,$$

that is the $\{d\}$ process is stationary, white, with zero mean and variance unity.

<u>Analyticity</u>

For each draw, $T_N(z)$ is a polynomial function of z ; in the large $N-$ limit its analyticity structure is given by the following theorem

*Almost surely, that is with probability 1, the unit circle $\mathcal{C}(0,1)$ is a natural boundary for $T_\infty(z)$.*

(PA's mimic natural boundaries as the support of a dense set of alternating zeros and poles).

<u>Probability law, moments and energy spectrum</u>

With the above assumptions on the $d$'s and for any given complex $z$, $T_N(z)$ is a random complex number.

To clarify this situation, we first specify <u>z to be real</u>. We note that the $\{d_m z^m\}$ are rrv, i, but not id since each has its own variance $z^{2m}$. Nevertheless the stability of the normal law under linear manipulations of rrv's implies that $T_N(z)$ is a gaussian variable with

$$\mathbb{E}[T_N(z)] = 0,$$

$$\mathbb{E}\left[T_N{}^2(z)\right] \;=\; 1 \;+\; z^2 \;+\; \dots \;+\; z^{2(N-1)} = \frac{1-z^{2N}}{1-z^2} \,.$$

So finally

$$T_N(z) = \sqrt{\frac{1-z^{2N}}{1-z^2}}\, g_N(z) \;,$$

where each $g_N(z)$ is a rrv distributed according to $\mathcal{N}(0,1)$. For $|z|$ equal or larger than 1, the large N-limit is meaningless. For $z \in\, ]-1,+1[$ , and for $N$ large enough, this equality takes the asymptotic form

$$T_N(z) \simeq \frac{1}{\sqrt{1-z^2}}\, g_N(z) \;;$$

thus, for $\rho = |z| \lesssim 1$ , the order of magnitude of the random z-Transform explodes like $2^{-1/2}(1-\rho)^{-1/2}$.

We now turn to the general case $\underline{z \in \mathbb{C} \setminus \mathbb{R}}$ . To give a precise meaning to *complex gaussian* random variable one usually resorts simply to its decomposition into its real and imaginary parts. A complex random variable

$$W = X + iY,$$

$$X \;=\; \mathcal{R}e(W),$$

$$Y \;=\; \mathcal{I}m(W),$$

is said gaussian if $X$ and $Y$ are real gaussian variables. The knowledge of the complex number $W$ is equivalent to the knowledge of the 2-dimensional real vector $\underline{W} = (X,Y)$ ; the probability distribution function (pdf) of the latter is, in the gaussian case, the 2-dimensional vectorial gaussian pdf

$$pdf(\underline{W}) = \frac{1}{2\pi}\frac{1}{\sqrt{det[\mathbf{K}]}} exp\left[-\frac{1}{2}\; \underline{W}\; \mathbf{K}^{-1}\,{}^t\, \underline{W}\right],$$

$$\mathbf{K} = \begin{bmatrix} \mathbb{E}\left[X^2\right] & \mathbb{E}\left[XY\right] \\ \mathbb{E}\left[XY\right] & \mathbb{E}\left[Y^2\right] \end{bmatrix},$$

(here in the zero mean case). Using this definition for $T_N(z)$, with non trivial $N \geq 2$, we have

$$X = \sum_{0 \leq m \leq N-1} \rho^m\, cos\, m\theta\, d_m \;,$$

$$Y \;=\; \sum_{0 \leq n \leq N-1} \rho^n\, sin\, n\theta\, d_n \;,$$

where $z$ has been written in the Argand decomposition

$$z \;=\; \rho\,(cos\,\theta \;+\; i\, sin\,\theta).$$

The elements of the autocorrelation matrix $\mathbf{K}$ thus read

$$\mathbb{E}\left[X^2\right] = \sum_{0 \le m \le N-1} \rho^{2m} \, cos^2 m\theta$$

$$\mathbb{E}\left[Y^2\right] = \sum_{0 \le n \le N-1} \rho^{2n} \, sin^2 n\theta$$

$$\mathbb{E}\left[XY\right] = \frac{1}{2} \sum_{0 \le l \le N-1} \rho^{2l} \, sin \, 2l\theta$$

and consequently

$$\Delta = det \, [\mathbf{K}] = \mathbb{E}\left[X^2\right] \mathbb{E}\left[Y^2\right] - \mathbb{E}\left[XY\right]^2 = \frac{1}{4} \frac{(1-\rho^{2N})^2}{(1-\rho^2)^2} - \frac{1}{4} \frac{1 - 2\rho^{2N} cos \, 2N\theta + \rho^{4N}}{1 - 2\rho^2 cos \, 2\theta + \rho^4}$$

If $\theta = 0(\pi)$, the determinant $\Delta$ vanishes; this is the already studied real case; the bi-gaussian collapses to a mono-gaussian. We note that if $\rho \ge 1$, the formalism cannot be extended to a $N = infinity$ régime. We thus consider now the case $\theta \ne 0(\pi)$, with $N \ge 2$ and $0 < \rho < 1$ ; determinant $\Delta$ can then be factorized according to

$$\Delta = \frac{\rho^{2N} sin^2\theta}{(1-\rho^2)^2 + 4\rho^2 sin^2\theta} \left[ \frac{(1-\rho^{2N})^2}{(1-\rho^2)^2} \frac{\rho^2}{\rho^{2N}} - \frac{sin^2 N\theta}{sin^2\theta} \right];$$

the first quantity in the bracket is always larger than $N^2$, while the second one is strictly positive and smaller than $N^2$. In this case the determinant is thus strictly positive; this inequality is the avatar of the Schwarz inequality in probability theory. We now concentrate on the asymptotic régime $\rho \lesssim 1$ and at the same time $\rho^{2N} \ll 1$, which is always possible if $N$ is large enough; from the last equality one can deduce that the determinant $\Delta$ then takes the asymptotic form

$$\Delta \simeq \frac{1}{16} \frac{1}{(1-\rho)^2} - \frac{1}{16} \frac{1}{sin^2\theta}.$$

This expansion breaks down in the vicinity of the curve $\rho = 1 - |sin\theta|$ plotted in polar coordinates in
Fig (5). Or, conversely stated, if $z$ is <u>near</u> the real axis, the explosive régime of $|T(z)|$ takes place only if, at the same time, $z$ is <u>very near</u> the unit circle and $N$ is <u>very large</u>. Standing outside this curve, the final form is

$$\Delta \simeq \frac{1}{16} \frac{1}{(1-\rho)^2} \quad .$$

In the same régime and under the same proviso, the autocorrelation $\mathbf{K}$ reads

$$\mathbf{K} \simeq \begin{bmatrix} \dfrac{1}{4} \dfrac{1}{(1-\rho)} & 0 \\ 0 & \dfrac{1}{4} \dfrac{1}{(1-\rho)} \end{bmatrix},$$

and the $pdf\left(\underline{W}\right)$ simplifies into

$$pdf\left(\underline{W}\right) \simeq \frac{1}{2\pi} 4(1-\rho) \, exp\left\{ -\frac{1}{2} 4 \, (1-\rho)(X^2 + Y^2) \right\}$$

which can be factorized using $pdf(X)$ and $pdf(Y)$. In other words the common order of magnitude of the rrv iid's $X$ and Y is $2^{-1}(1-\rho)^{-\frac{1}{2}}$.

The picture which emerges for a typical draw of $T(z)$ for white gaussian data and complex $z$ is thus as follows. On the unit circle $\mathcal{C}(0,1)$, there is a natural boundary. On any inner circle of radius $\rho = |z| \lesssim 1$, the modulus of $T(z)$ scales like $(1-\rho)^{-\frac{1}{2}}$. However the $|T(z)|$ explosion region dies off as one approaches the real domain. As we will see in §14, the Padé Filtering captures strikingly well this picture: the density of random poles is sizeable near $\mathcal{C}(0,1)$, except in the complex vicinity of $\pm 1$, where the singularity is milder. Exactly on the real axis in turn, the scaling of the modulus of $T(z)$ is valid again and the density of poles of the PA's is unidimensional; one recovers the global attractive power of the real axis acting on the roots of real polynomials (here the denominator of the PA) we met in §4. Finally, in §8, we will extend the study to some colored (= not white) gaussian stationary processes.

Figure 5

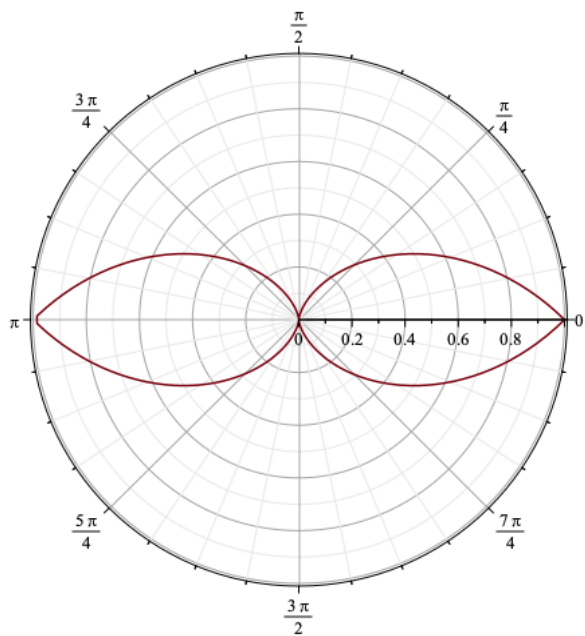

Finally the mathematical expectation of the analytic extension of the energy spectrum reads

$$\mathbb{E}[\mathcal{E}_N(z)] = \frac{1}{N} \sum_{0 \le m,\ l \le N-1} \mathbb{E}[d_m\, d_l]\, z^{m-l}.$$

Hence, using the whiteness and the $\mathcal{N}(0,1)$ pdf of the $d_m$'s,

$$\mathbb{E}[\mathcal{E}_N(z)] = 1.$$

### b. The random autocorrelation and the Wiener-Khintchine theorem

To a random process can be associated its various unequal times moments. We concentrate here on $\mathbb{E}[X_m X_l]$, which is a function of $m$ and $l$. If the process is statistically stationary, this 2-points moment depends only on $k = l - m$. We consider its estimate

$$C_N(k) = \frac{1}{N} \sum_{0 \le m \le N-1-k} X(m)\, X(m+k),$$

$$0 \le k \le N-1.$$

It is itself a random process of length $N$ and we will call it the "random autocorrelation". It contains some of the physics of the original process $\{X\}$. Its successive values may be used to generate another random $z$-Transform, thus having available two $z$-Transforms associated to the original process, namely

$$T_N^X(z) = \sum_{0 \le n \le N-1} X(n) z^n \,,$$

and

$$T_N^C(z) = \sum_{0 \le k \le N-1} C_N(k) z^k .$$

It is a natural issue to compare the physical and analytical content of the two. For finite $N$, one can formally derive the remarkable identity

$$T_N^C(z) + T_N^C\left(\frac{1}{z}\right) - T_N^C(0) \;=\; \frac{1}{N}\, T_N^X(z)\, T_N^X\left(\frac{1}{z}\right) = \mathcal{E}_N^X(z) \,.$$

The corresponding equality holds in the $N$ goes to infinity limit; in this case, it is known as the Wiener- Khintchine theorem. In this limit, both $T^X(z)$ and $T^C(z)$ have zeros and singularities, but the theorem does not provide any unambiguous link between their respective analytic structures. To explore this link, directly and using Padé filtering, we have to resort to examples. Here we briefly examine the random autocorrelation and its random $z$-Transform for the white gaussian large length process.

So we consider the collection $\{X_m\}_{m=0}^{N-1}$, where the X's are rrv iid according to the gaussian law $\mathcal{N}(0,1)$. First we note that the product entering the definition of $C_N(k)$ are not gaussian and not independent; nevertheless, a strong version of the Central Limit Theorem ensures that the sum of these products becomes gaussian in the large $N$ régime. After calculation of the mean and variance of the C's one concludes precisely that

$$C_N(0) = 1 + \frac{\sqrt{2}}{\sqrt{N}} \xi(0),$$

$$C_N(k) = \frac{1}{\sqrt{N}} \sqrt{1 - \frac{k}{N}}\, \xi(k) \,, \qquad for\ k > 0,$$

where the $\xi$'s are rrv which for large $N$ become iid $\mathcal{N}(0,1)$. For infinite $N$, the ergodic limit $C_\infty(k)$ reaches $\mathbb{E}[X_l X_{l+k}] = \delta_{k,0}$. Using the definition of the z-Transform of the C's, one notes that, for given z and large $N$, $T_N^C(z)$ is an entire series with gaussian coefficients; apart from the peculiar mean and variance of $C_N(0)$, those coefficients comply with the assumptions concerning the d's data envisaged in the previous subparagraph ; so the conclusions convey to the present case, especially the existence of a natural boundary on the unit circle for $T_\infty^C(z)$. Specifying to real z with $-1 < z < 1$ and large $N$, one obtains

$$\mathbb{E}\big[T_N^C(z)\big] \;\simeq\; 1 \,,$$

and

$$\mathbb{E}\left[\big[T_N^C(z) - 1\big]^2\right] \simeq \frac{1}{N}\left[1 + \frac{1 - z^{2N}}{1 - z^2}\right] - \frac{1}{N^2} z^2 \frac{1 - N z^{2N-2} + (N-1) z^{2N}}{(1 - z^2)^2} \,;$$

thus for $T_N^C(z)$ the random expression

$$T_N^C(z) \simeq 1 + \frac{1}{N} \frac{\sqrt{2N - (3N+1)z^2 + Nz^4 + z^{2N+2}}}{|1-z||1+z|} h_N(z),$$

where the $h_N(z)$ are rrv iid $\mathcal{N}(0,1)$. We now turn to the sub-régime where $z$ is closed to $1$ and $N$ large enough to ensure that $z^{2N}$ is very small. It yields

$$T_N^C(z) \simeq 1 + \frac{1}{\sqrt{N}} \frac{1}{\sqrt{2}} \frac{1}{\sqrt{1-z}} h_N(z),$$

to compare with

$$T_N^X(z) \simeq \frac{1}{\sqrt{2}} \frac{1}{\sqrt{1-z}} g_N(z).$$

The quasi-singularity is the same, but for an $1/\sqrt{N}$ amplitude factor.

# PART B

# MEROMORPHIC TOOLS IN A NOISY SETTING

IV.  Polynomial and rational tools, old and new

    6.  Resonances

    7.  Roots of random generalized polynomials

    8.  ARMA generated noises. Working out the AR(1) and AR(2) cases

    9.  Szegö polynomials, deterministic and random.
           a.  Spectral identification via deterministic Szegö polynomials
           b.  Random Szegö polynomials

V. Rational approximants with noise

    10. Noisy Rational Interpolants: toy model and general rational case

    11. Noisy Padé Approximants: unique simple pole, general rational case, generalization

# PART B: MEROMORPHIC TOOLS IN A NOISY SETTING

## IV Polynomial and rational tools, old and new

### 6. Resonances

Resonances, viewed as poles of functions or operators, are a classical and contemporary research topic in natural sciences and technology. One is often interested in their motion in the complex as some parameters vary. Such phenomena are embodied in the behavior of the roots of a polynomial as a function of the coefficients. We note that rational and polynomial functions appear also in approximation and representation theory. In all these fields one is thus interested in the zeros of deterministic polynomials (or entire functions). What if noise comes in? In this part we examine some answers by presenting a few selected prototypical examples, which are all somehow related to the Wiener filter transfer function.

### 7. Roots of random polynomials and random generalized polynomials

We consider two extreme examples of random finite polynomials. In the first one, a large signal to noise ratio is assumed. Precisely one considers

$$P_n(z) = \Lambda z^n + \sum_{k=0}^{n-1} a_k z^k$$

where the $a_k$'s are rrv iid $\mathcal{N}(0,1)$ and $\Lambda$ is a certain (not random) positive 'large' number. Numerical simulations indicate a concentration of the random roots near a circle inside the unit circle $\mathcal{C}(0,1)$, suggesting the rescaling of the free variable $z$. In posing $z = \Lambda^{\frac{-1}{n}} y$ we get

$$P_n[z(y)] = y^n + a_0 + \sum_{k=1}^{n-1} a_k y^k \;;$$

the first two terms of the rhs are $\mathcal{O}(1)$; the others contribute at most for a global $\mathcal{O}(\Lambda^{\frac{-1}{n}})$ noise and we neglect them in a first order very low disorder expansion. One is left with the model problem

$$y^n + a_0 = 0 \,.$$

In a given realization, the solutions of this model sit on $n$ rays among the $2n$ rays in the complex emanating from the origin with the arguments $2\pi \frac{l}{n}, \, l = 0, \dots, \, n-1$ and $\frac{\pi}{n} + 2\pi \frac{m}{n}, m = 0, \dots, n-1$; the mean of their modulus is readily obtained as the $\left(\frac{1}{n}\right)$–th moment of the gaussian law, namely

$$\mathbb{E}\left[|y|\right] = 2^{\frac{1}{2n}} \frac{1}{\sqrt{\pi}} \, \Gamma\left(\frac{n+1}{2n}\right).$$

The $n$ random solutions of the $z$ initial problem are thus fluctuating around the 2n nodes of a quasicrystal, located on a circle of radius proportional to $\left(\frac{\sqrt{2}}{\Lambda}\right)^{\frac{1}{n}}$ for $\Lambda$ and $n$ large; the latter quantity is close to 1 in the sub-régime $n \gg \ln \Lambda$. See figures 7.1 and 7.2 displaying the quasicrystal and the scaling law of its radius for some values of $n$ and $\Lambda$.

The second example is the $\Lambda = 0$ case, displaying extreme disorder and called "homogeneous" or "Kac" polynomials. In §4, we have explored the real roots of such polynomials. Properly adapted to complex random roots, the same formalism yields a generalization of the Kac's formula for the 2D density of random roots; in the large $n$ régime, the latter are uniformly distributed on a circle close to $\mathcal{C}(0,1)$, but for the neighborhood of the real axis. There is no crystal here as can be seen on figure 7.3.

Figure 7.1

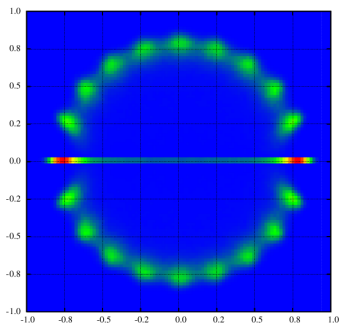

Figure 7.2

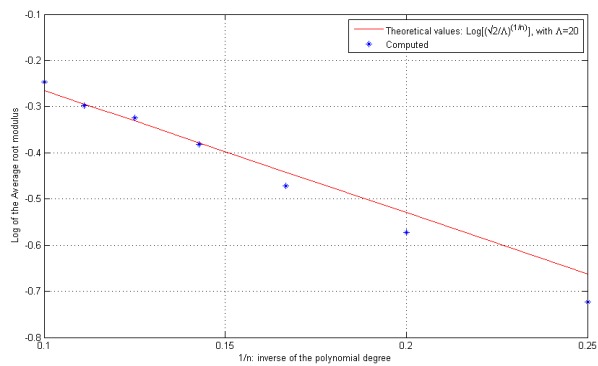

Figure 7.3

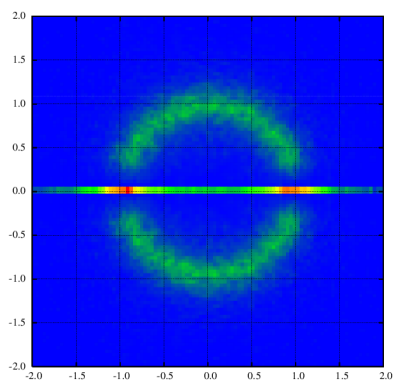

These studies have been extended by various authors to 'generalized polynomials'

$$P_n(z) = \Lambda\Phi(z) + \sum_{k=0}^{n-1} a_k f_k(z),$$

where $\Phi$ and the $f_k$'s are analytic functions; also restrictions like gaussianity of the $a_k$'s have been lifted, and orthogonal $f_k$'s with respect to $\mathcal{C}(0,1)$ (see §9) or with respect to another line $\Gamma$ have been envisaged; in the latter case the zeros are equidistributed with respect to the equilibrium measure supported by $\Gamma$. Applications include signal processing as well as string theory.

## 8. ARMA generated noises. Working out the AR(1) and AR(2) cases

Consider the difference equation for the random process $\{X(m)\}_{m=0}^{N-1}$

$$X(m) + a_1 X(m-1) + \cdots + a_P X(m-P) = \Phi(m) + b_1 \Phi(m-1) + \cdots + b_Q \Phi(m-Q)$$

where $\{\Phi(l)\}_{l=0}^{N-1}$ is a random stationary white gaussian source process and $\{\Phi\}$ and $\{X\}$ both obey causal boundary conditions

$$l < 0 \;\Rightarrow\; \Phi(l) = 0,$$

$$m < 0 \;\Rightarrow\; \mathrm{X}(m) = 0.$$

For the sake of stability and definiteness, the $z$-polynomial built with the $a's$ coefficients has its zeros outside the unit disk in the complex; it does not have any zero in common with the polynomial built with the $b's$. Disregarding the very first times, the $\{\mathrm{X}\}$ random process is stationary; due to the linearity of the equation it is itself gaussian but it is not white. The l.h.s. of the equation is an Auto Regressive linear combination of the X's, of degree $P$, or AR($P$); the r.h.s. is a Moving Average linear combination of the $\Phi$'s, of degree $Q$, or MA($Q$). The solution $\{X(m)\}_{m=0}^{N-1}$ is called an ARMA $(P, Q)$ process. Such processes are widely used to model actual colored gaussian data. Here we use them the other way around, as theoretical or numerical generators of colored gaussian processes. In this paragraph we develop the theme announced in §1 and §3, that is the contradistinction between the complex poles of the $z$-Transform of the process and the real peaks of its Fourier energy spectrum; this topic is studied on the AR(1) and AR(2) cases which we completely work out.

<u>AR (1)</u> In this case, the difference equation reduces to

$$X(m) \,-\, a\, X(m-1) \;=\; \Phi(m),$$

with $0 < a < 1$. The z-Transforms $T_N^\phi(z)$, of the $\phi's$, and $T_N^X(z)$, of the X's, are related by the recurrent equality

$$T_N^X(z) \,-\, a\, z\, T_{N-1}^X(z) \;=\; T_N^\phi(z)\,;$$

hence for the infinite $N$ limit

$$T_\infty^X(z) \;=\; \frac{1}{1 - az} T_\infty^\phi(z)$$

and

$$\mathcal{E}_\infty^X(z) = \frac{z}{(1 - az)(z - 1)} \mathcal{E}_\infty^\phi(z).$$

Taking the expectation value of both sides of the last equality, one gets

$$\mathbb{E}[\mathcal{E}_\infty^X(\theta)] = -\frac{1}{a} \frac{z}{z^2 - \left(a + \frac{1}{a}\right)z + 1} = -\frac{1}{2}\frac{1}{a}\frac{1}{cos\theta - cosh(lna)},$$

$$z \;=\; e^{i\theta}, -\pi \le \theta \le \pi.$$

The analyticity structure of $T_N^\phi(z)$ has been discussed in §5 ; we conclude that $T_\infty^X(z)$ has almost surely a natural boundary on $\mathcal{C}(0,1)$ and an exterior pole at $z_* = 1/a$. The extrema of $\mathbb{E}[\mathcal{E}_\infty^X(\theta)]$ are governed by the condition $sin\,\theta = 0$ and are located at $-\pi\,;0\,;\pi$; the finite peak is located at $\theta_c \;=\; 0$ see figure 8.1.

Figure 8.1

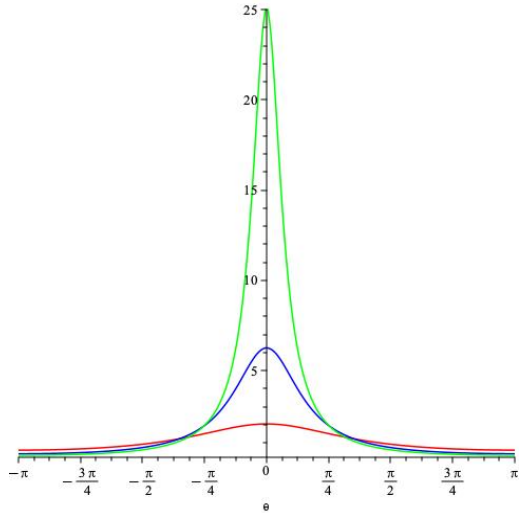

In the AR(1) models, the energy is concentrated in the large scales, that is small $|\theta|$, hence the name pink noise ; the location of the peak does not vary as $a$ varies. In contrast the analyticity properties of $T(z)$ give a richer and $a$-dependent picture; the natural boundary of $T(z)$ accounts for the randomness and its pole is a marker of the reddish character of the process.

<u>AR (2)</u> In this case the difference equation may be conveniently written

$$X(m) - 2\frac{1}{r}\cos\varphi\, X(m-1) + \frac{1}{r^2}X(m-2) = \Phi(m),$$
$$0 \leq m \leq N-1, \quad 1 < r, \quad -\pi \leq \varphi \leq \pi,$$

without loss of generality. Using the definition of the z-Transforms of the X's and the $\Phi$'s, one derives

$$T_\infty^X(z) \;=\; \frac{r^2}{z^2 - 2r\cos\varphi\; z\; +\; r^2}\; T_\infty^\phi(z).$$

From the results of §5 concerning $T_\infty^\phi(z)$, we conclude that $T_\infty^X(z)$ has a natural boundary on the unit circle $\mathcal{C}(0,1)$ in the complex; in addition, it has a pair of complex conjugate exterior poles located at $r\,exp(\pm\,i\,\varphi)$.

Turning to the Fourier energy spectrum

$$\mathcal{E}_N^X(z) = \frac{1}{N}\,\mathrm{T}_N^X(z)\,\,\mathrm{T}_N^X\left(\frac{1}{z}\right),$$

we obtain its expectation value in the large N limit

$$ES\,(\theta) \;\equiv\; \mathbb{E}\left[\mathcal{E}_\infty^X(z)\right] \;=\; \frac{r^4}{4\,r^2 cos^2\,\theta \;-\; 8\,r^2 q\,cos\,\theta \;+\; (1-r^2)^2 \;+\; 4\,r^2\,cos^2\varphi}\,,$$

where

$$z = exp(i\theta), \qquad -\pi \le \theta \le \pi,$$

$$q = \frac{1}{2}\left(r + \frac{1}{r}\right)\cos\varphi.$$

The extrema of $ES(\theta)$ are controlled by the zeros of its derivative with respect to $\theta$, that is by the equation

$$\sin\theta\,[\cos\theta - q] = 0,$$

whose solutions are zero (modulo $\pi$) and $\theta_*$ such that

$$\cos\theta_* = q\,.$$

The latter equality is meaningful only if $q \in [-1, +1]$, a constraint which can be mapped into the inequality

$$|\cos\varphi|\,r^2 - 2\,r + |\cos\varphi| \le 0;$$

the solution of this inequality is in turn

$$r \le r_+,$$
$$r_+ = \frac{1 + |\sin\varphi|}{|\cos\varphi|}.$$

See in figure 8.2 the plot of $r_+$ as a function of $\varphi$; if $r$ is larger than $r_+$ there is no solution $\theta_*$; if $r$ is between unity and $r_+$, $\theta_*$ is such that

$$\cos\theta_* = \frac{1}{2}\left(r + \frac{1}{r}\right)\cos\varphi = \cosh(lnr)\,\cos\varphi.$$

Figure 8.2

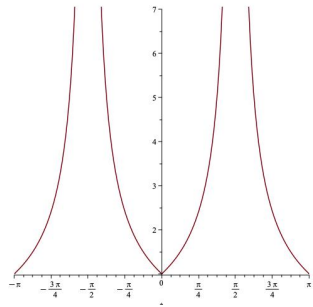

The following scenario thus emerges for the energy spectrum of this model as its parameters $r$ and $\varphi$ vary. See figures 8.3.

For given $\varphi$

| | |
|---|---|
| $r_+(\varphi) < r$ | one peak at the origin<br>one minimum at $\theta = \pm\pi$ |
| $r = r_+(\varphi)$ | a flat peak at the origin<br>one minimum at $\theta = \pm\pi$ |
| $1 < r < r_+(\varphi)$ | one minimum at the origin<br>two peaks at $= \pm\theta_*$<br>one minimum at $\theta = \pm\pi$ |

One notes that

| | |
|---|---|
| $\varphi = 0\ (\pi)$ | the double peak never exists |
| $\varphi = \pm\dfrac{\pi}{2}$ | the double peak always exists<br>$\theta_* = \dfrac{\pi}{2}$ |

When the double peak is present it is located at $\pm\theta_*$ such that $|\theta_*|$ is always smaller than $|\varphi|$ (if $\varphi = \pm\frac{\pi}{2}$, equality is reached). We comment that the motives drawn by the energy spectrum vary in an intricate way and are almost never a faithful indication of the analyticity structure of the z-Transform. In part C we shall see how the PA's properly capture the latter structure even with a finite number N of data. Thus our motto

*"Better study the complex poles of the z-Transform of a process rather than the real peaks of its energy spectrum."*

Figure 8.3

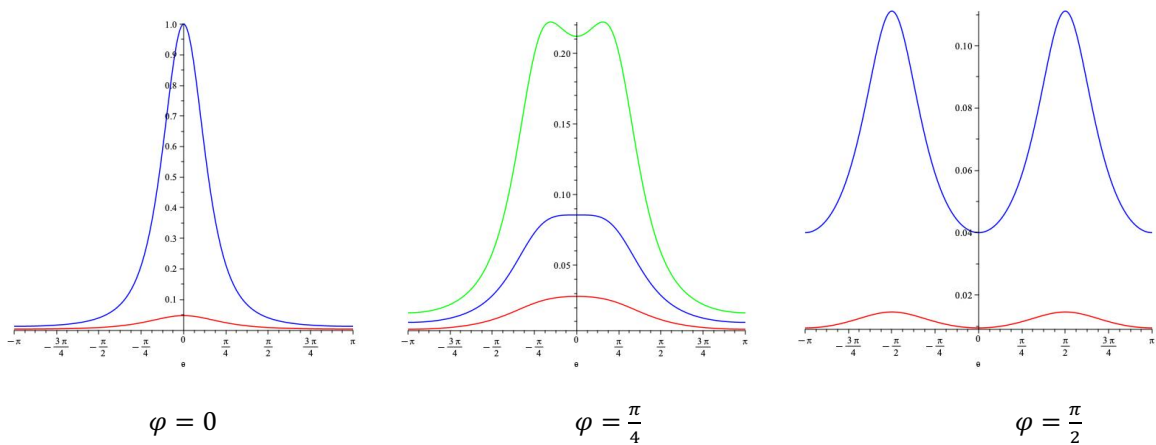

$\varphi = 0$ $\qquad\qquad\qquad\qquad \varphi = \dfrac{\pi}{4}$ $\qquad\qquad\qquad\qquad \varphi = \dfrac{\pi}{2}$

9. Szegö polynomials, deterministic and random.

   a. Spectral identification via deterministic Szegö polynomials.

We assume that the reader is familiar with the classical theory of orthogonal polynomials on the line. We recall here only that orthogonality is usually defined using the inner product of two functions $f$ and $g$

$$< f, g > = \int_I f(x)g(x)d\mu(x)$$

where $\mu(x)$ is a positive measure and $I$ is a finite or infinite interval on $\mathbb{R}$. Given a family $\{P_n(x)\}$ of polynomials orthogonal with respect to this product, their zeros exhibit many properties including interlacing and localization of the nodes of the Gauss quadrature method. We now concentrate on the complex case introduced by G. Szegö in using the unit circle $\mathcal{C}(0,1)$ in $\mathbb{C}$ instead of the interval $I$ on $\mathbb{R}$. More precisely one considers the inner product between two real-analytic functions of the variable z defined by

$$< f, g > = \int_{-\pi}^{+\pi} f(e^{i\theta})\overline{g(e^{i\theta})}\ \rho(\theta)d\theta\ ,$$

where $\rho(\theta)$ is the positive density of a measure and the bar denotes complex conjugation. The family $\{S_p(z)\}$ of Szegö polynomials orthogonal with respect to this inner product is prescribed by

$$< S_p(z), S_r(z) > \propto \delta_{p,r}\ ,$$

where the $S_p$'s are monic polynomials in z. The connection with data analysis is provided with the additional assumption that $\rho(\theta)$ is the Fourier energy spectrum of a discrete-time signal. Suppose now that this signal is the deterministic sum of damped cosine functions $\mathcal{A}_j e^{-\lambda_j t}\cos(\omega_j t + \varphi_j), j = 1, \dots, \mathcal{U}$ and consider the associated Szegö polynomials $S_p(z)$ with a degree $p$ larger than 2 $\mathcal{U}$. Numerical experiments and rigorous estimates point to the following scenario: as $p$ and the length $N$ of the signal increase, 2 $\mathcal{U}$ roots of $S_p(z)$ tend towards the 2 $\mathcal{U}$ resonances, $\Omega_j = \exp(-\lambda_j\tau + i\omega_j\tau)$, $\overline{\Omega_j} = \exp(-\lambda_j\tau - i\omega_j\tau)$, where $\tau$ is the discrete-time step; the $(p - 2\ \mathcal{U})$ remaining roots tend towards limiting points whose modulus is smaller than the moduli of the resonances; thus a selective detection of the (possibly damped) trigonometric components of the signal by the zeros of the associated Szegö polynomials.

   b. Random Szegö polynomials

What if noise comes in? The Fourier energy spectrum of the signal is now itself a random quantity and the roots of the associated Szegö polynomials are themselves random, which at first sight may look a hopeless situation. In fact it is quite the opposite: as already noticed, randomness creates order, at least in some asymptotic situations. Let us consider a not white process and take $N$ and $p$ large; then small statistical clouds of zeros tend towards the resonances of the system; the remaining zeros, called noisy zeros, concentrate on an anulus inside $\mathcal{C}(0,1)$. In the regime $p \ll lnN$, the radius of this anulus is small; in the regime $1 \ll \ln N \ll p \ll N$, the radius of this anulus is close to 1, but its thickness is very small; in both cases there is a clear separation between resonances and noisy zeros. We bring to the attention of the reader the existence of links between Szegö polynomials, Padé approximants, Wiener Transfer functions and denominators of AR systems energy spectra.

# V. Rational approximants with noise

## 10. Noisy Rational Interpolants: toy model and general rational case

Suppose a function $\varphi(z)$ is known at certain nodes $z_j$, $j = 1 \ldots, M$; those nodes are real but not necessarily regularly spaced. One looks for an interpolating function $f_\varepsilon(z)$ such that

$$f_\varepsilon(z_j) = \varphi(z_j)(1 + \varepsilon r_j), \; j = 1 \ldots, M,$$

with $f_\varepsilon(z)$ belonging to a certain functional class; $\varepsilon$ is a – eventually vanishing – positive parameter and the $r_j's$ are rrd, iid. We will stick with the rational class, assuming

$$f_\varepsilon(z) = \frac{P_p(\varepsilon; z)}{Q_q(\varepsilon; z)},$$

where $P_p$ and $Q_q$ are finite polynomials in $z$; $f_\varepsilon(z)$ is called a Rational Interpolant (RI) or a Padé Approximant of type II. One is interested in the influence of $\varepsilon$ and of the statistics of the $r_j's$ on the statistics of the roots of $P_p$ and $Q_q$. A first numerical exploration with the example

$$\varphi(z) = tanh(z)$$

reveals a phenomenon akin to the Froissart phenomenon known for PA's: some poles and zeros of the RI reflect the analytic structure of $\varphi(z)$ while some come in doublets localized inside the interpolation interval $[z_1, z_M]$. See Figure 10.

Figure 10

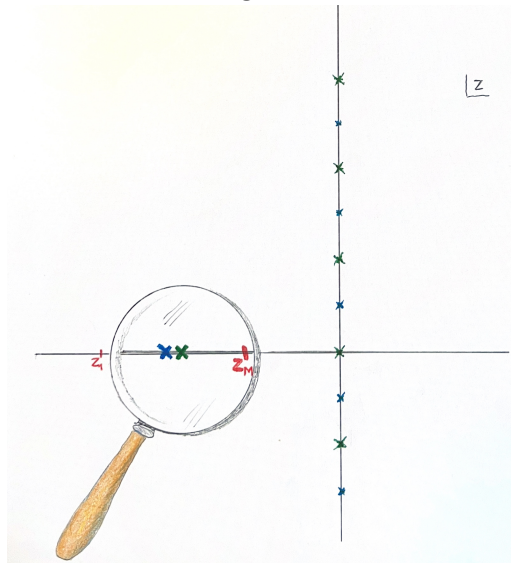

In the sequel of this paragraph we study this phenomenon for $\varphi(z)$ rational,

$$\varphi(z) = \frac{T_m(z)}{B_n(z)}.$$

**Toy model.** To best look at the steps of the calculation, we resort to the constant $\varphi(z)$ toy model: $m = n = 0$, $\varphi(z) \equiv \phi$. We also set $p = q = 1$ and $M = 3$. The interpolation constraints then read

$$P_1(z_j) = \phi\,(1 + \varepsilon r_j)\,Q_1(z_j), \qquad j = 1,2,3.$$

Without loss of generality the polynomials $P_1(z)$ and $Q_1(z)$ may be written

$$P_1(z) = p_0 + p_1 z,$$

$$Q_1(z) = 1 + q_1 z.$$

The nodes are denoted

$$z_1 = g - 1, \qquad z_2 = g + e, \qquad z_3 = g + 1,$$

allowing for irregular spacing. Finally, the set of equations is a 3x3 inhomogeneous linear system for $p_0$, $p_1$, $q_1$, namely

$$p_0 + (g - 1) p_1 - (1 + \varepsilon r_1)(g - 1) \phi q_1 = (1 + \varepsilon r_1)\phi,$$

$$p_0 + (g + e) p_1 - (1 + \varepsilon r_2)(g + e) \phi q_1 = (1 + \varepsilon r_2)\phi,$$

$$p_0 + (g + 1) p_1 - (1 + \varepsilon r_3)(g + 1) \phi q_1 = (1 + \varepsilon r_3)\phi.$$

The interpolating function takes thus the form`

$$f_\varepsilon(z) = \frac{K_1(z) + \varepsilon \overline{P_1(z)}}{K_1(z)} \phi \; ;$$

$\overline{P_1}(z)$ is independent of $\varepsilon$ and $K_1(z)$ reads

$$K_1(z) = (r_3 - r_1) + e(r_3 - 2r_2 + r_1) - [(1 - e)r_1 - 2 r_2 + (1 + e)r_3](z - g).$$

In the small $\varepsilon$ régime, in each realization, the root of the denominator of $f_\varepsilon(z)$ is close to the root of its numerator, which proves the existence of a Froissart type doublet for the studied noisy RI. The random polynomial $K_1(z)$ is called the Froissart polynomial (FP); it governs the statistics of the position of the (real) doublet. The probability distribution of the position, with respect to $g$, of this doublet, is given by

$$\rho(\xi) = \mathbb{E}\left[\delta(\xi - \xi_*)\right]$$

under the constraint

$$K_1(z_*) = 0,$$

$$z_* = g + \xi_*.$$

Assuming a gaussian $\mathcal{N}(0,1)$ probability law for the $r_j$'s, the doublet density can be calculated exactly and reads

$$\rho(\xi) = \frac{1}{\pi} \frac{1}{\Lambda} \frac{1}{1 + (\xi - \xi_c)^2/\Lambda^2},$$

$$\xi_c = e \frac{4}{3} \frac{1}{1 + (e/\sqrt{3})^2}$$

$$\Lambda^2 = \frac{1}{\sqrt{3}} \frac{(1 - e^2)}{1 + (e/\sqrt{3})^2},$$

$$-1 \leq e \leq 1.$$

This Cauchy-Lorentz law can be summed which yields in particular

$$\mathcal{P}\left[z_c \in [z_1, z_3]\right] = \frac{2}{3}.$$

Independently of the central node displacement $e$, the large majority of the doublets lie inside the interpolation interval; this proves the condensation of the doublets, the second aspect of the Froissart phenomenon; in the present RI case, the condensation takes place in the vicinity of the interpolation interval.

    <u>General rational case.</u> Turning to the $[m+k \,/ n+k]$ RI to a noisy rational $[m \,/n]$ function, the interpolation constraint again gives rise to a linear algebraic $(m+n+2k+1)\mathrm{x}(m+n+2k+1)$ system for the coefficients of the numerator and denominator of $f_\varepsilon(z)$. Using formal manipulations on the corresponding Cramer determinants, one can prove the formula

$$f_\varepsilon(z) = \frac{T_m(z)K_k(z) + \varepsilon \; \bar{P}_{m+k}(\varepsilon; z)}{B_n(z)K_k(z) + \varepsilon \; \bar{Q}_{n+k}(\varepsilon; z)};$$

its structure again ensures the generalization of the first aspect of the Froissart phenomenon, namely the existence of small elongation zero-pole doublets in the small $\varepsilon$ régime. Assuming gaussian $r_j's$ and $k = 1$,
the probability distribution of the (real) doublet is again a Cauchy law and its summation yields

$$\mathcal{P}\left[z_c \in [z_1, z_M]\right] \geq \frac{1}{2} \,,$$

ensuring the Froissart condensation in the vicinity of the interpolation interval.

11. Noisy Padé Approximants: unique simple pole, general rational case, generalization.

    We now turn to classical Padé Approximants (PA) to noisy functions. Suppose a function $\varphi(z)$ is known through its Taylor coefficients $\{c_j\}$. One considers the perturbed function $\varphi_\varepsilon(z)$ according to

$$\varphi_\varepsilon(z) = \sum_{j \geq 0} (c_j + \varepsilon \; r_j) z^j \,,$$

where $\varepsilon \in \mathbb{R}_+$ and the $r_j's$ are rrv iid. One looks for a rational function

$$f_\varepsilon(z) = \frac{P_p(z)}{Q_q(z)}$$

such that $f_\varepsilon(z)$ is 'equal' to $\varphi_\varepsilon(z)$ up to a certain order in the z-expansion. More precisely one looks for

$$P_p(z) = \sum_{j=0}^{p} p_j \, z^j \,,$$

$$Q_q(z) = 1 + \sum_{l=1}^{q} q_l \, z^l \,,$$

such that

$$P_p(z) - \varphi_\varepsilon(z)Q_q(z) = 0$$

up to the order $(p + q + 1)$. One is interested in the influence of $\varepsilon$ and of the $r_j's$ on the statistics of the roots of $P$ and $Q$. In the sequel of this paragraph we stick with $\varphi(z)$ rational.

<u>Unique simple pole</u>. Here we consider $\varphi(z) = 1/(1 - z)$ , whose Taylor series is such that $c_j = 1, \forall j.$ One first examines the $\left[ n - 1/n \right]$ PA's to this geometric series. As in the RI case, one notices that, given $z$ and the $r_j$'s, the governing constraint is a linear system for the $p_l$'s and the $q_m$'s which is solved using Cramer determinants. A combination of elementary manipulation of those determinants allows one to prove the formula

$$\left[ n - 1/n \right](\varepsilon; z) = \frac{K_{n-1}(z) + \varepsilon \ \bar{P}_{n-1}(\varepsilon; z)}{(1 - z)K_k(z) + \varepsilon \ \bar{Q}_n(\varepsilon; z)}$$

where $K_{n-1}, \bar{P}_{n-1}, \bar{Q}_n$ are polynomials in $z$. In the small $\varepsilon$ régime, this formula embodies the existence of a pole near $+1$ and the existence of $(n - 1)$ Froissart doublets of small extension located near the roots of $K_{n-1}(z)$.

<u>General Rational Case</u>. Following the same guidelines, the Froissart formula can be extended to the general rational $\varphi(z)$ case.

<u>Generalization</u>. In a further generalization, one considers a sort of combination of RI and PA called Multiple Points Padé Approximants (MPPA), where the approximation is written simultaneously in terms of the Taylor expansions around various nodes: again a Froissart formula can be derived.

PART C

PADÉ FILTERING

VI. Padé filtering: principles
　　　12. First works and extensions
　　　13. Proof of concept
　　　14. Prototypical analyticity patterns: $T(z)$ and PA's to $T(z)$

VII. Padé filtering in the making
　　　15.Testing on signals that are expected from the Gravitational Universe
　　　16.Testing on Virgo output channels subject to especially environmental noises

VIII. Further
　　　17. Actual implementation. Generalizations. The Froissart debate.

# PART C: PADÉ FILTERING

## VI. Padé filtering: principles

### 12. First works and extensions

In a precursory paper published in 1996, D. Bessis demonstrated the potential of PA's in noise filtering (JCAM, <u>66</u>, 1996, 85-88). The example was a set of electron-atom scattering data, for which the cross-section $\sigma(t)$ is an analytic function of $t$, one of the collision parameters; this function is perturbed by experimental errors. Assuming analyticity for the perturbed function, one explores it using PA's of type II, that is (noisy) RI's. The crucial fact is that the distributions of the poles and zeros of the latter split into classes reflecting the experimental noise, the computing noise and the analytic structure of the unperturbed original function $\sigma(t)$. The long cherished hope to separate message and noise, even indeed to erase the latter, seemed at hand, at the only cost to adopt the setting of complex analysis. The suggested mechanism is the mimicking of the natural boundary of the noisy $\sigma(t)$ by Froissart doublets; the net product is data compression, since in the example, 14 data are eventually represented using a rational function with only 3 parameters. The mathematical foundations and the applicability raised as many issues which were tackled later on by many authors; here we stress the work of P. Barone (Roma), D. Bessis and coworkers (Houston) and the Nice-Warsaw collaboration. Before reviewing such contributions, we recall the two main steps of Padé Filtering (PF), as announced in §1: (i) to characterize a set of, possibly noisy, $N$ data by the analytic structure of its $z$-transform; (ii) to study the latter using Rational Approximants. From the beginning one of the motivations was the application to interferometric GW detection, where the noise to signal ratio is gigantic. A proof of concept in this setting is given in the next paragraph (§13). Here we proceed with a few theoretical advances.

To apply the method to a general random process, it is relevant to first understand its outcome in paradigmatic cases. This was done by P. Barone (2005) who studied the purely noisy complex white gaussian case; he has shown that, for large data number $N$ and degree $n$ of the denominator of the PA of the $z$-transform of the data, the poles of the PA's "condense" uniformly on the unit circle. The argument relies on properties of the logarithmic potential of the density of poles in the complex; a workable model of the latter quantity was also proposed in the same paper. The rotational invariance is broken if the process (i) is not white as numerically explored[§§] by JDF and coworkers (see Fig. 14.d); or (ii) if it contains a deterministic component as stressed by the Houston group. Also, the concept of Froissart doublets is valid for the type II PA's and for the mix of type I and type II called MPPA (see §10 and 11); this was proven by M. Pindor and JDF in devising a determinantal formalism akin to the one developed by J. Gilewicz and M. Pindor for the type I PA.

### 13. Proof of concept

We formulate this paragraph in the setting of an apparatus in a stationary régime with many stages and linearly subjected to many inputs, random or deterministic; we also assume that between two successive stages, the state of the apparatus undergoes linear evolution; finally the state of the apparatus is collected at each stage via a linear filter, giving rise to many outputs. The functioning of any GW interferometer detector like Virgo can be modeled this way.

---

[§§] It would be valuable to have a rigorous proof e.g. for ARMA Processes.

We denote $\mathcal{I}_j$ the input at stage $j$, subject to convolution by $\mathcal{M}_j$ ; we denote $S_j$ the state at stage $j$, which evolves into $S_{j+1}$ under convolution by $\mathcal{E}_j$ ; we denote $O_j$ the output at stage $j$, obtained in convoluting $S_j$ by $\mathcal{L}_j$. Going back to the language of §1, we observe that the classical Fourier-convolution theorem holds for DFT and thus extends to the $z$-Transform. Expressing it for the stage $(j+1)$ and $(j)$ of the above apparatus, we get

$$T^{O_{j+1}} = T^{\mathcal{E}_j}\frac{T^{\mathcal{L}_{j+1}}}{T^{\mathcal{L}_j}}T^{O_j} + T^{\mathcal{E}_j}\,T^{\mathcal{L}_{j+1}}\,T^{\mathcal{M}_j}\,T^{\mathcal{I}_j}.$$

The analytic structures of these various z-Transforms are thus related. We point out that the singularities of $T^{O_j}(z)$, for each random realization, have to reappear among the singularities of $T^{O_{j+1}}(z)$ for the same realization. In a random case, the Padé Filtering method delivers histograms of the zero/pole positions on $\mathbb{C}$. Our statement then translates into "the local maxima of the polar histogram of $T^{O_j}(z)$ have to reappear among the local maxima of the polar histogram of $T^{O_{j+1}}(z)$". We stress that no equivalent statement does exist for the real peaks of the Fourier energy spectra of the successive output channels. The statement is successfully depicted on the two following figures which show the result of the Padé Filtering applied to two different signal sampling output channels of Virgo V1:Pr-B7-DC (photodiode placed at the very end of the North arm) and the V1:Pr-B1p-DC (photodiode placed at the dark port).

Figure 13

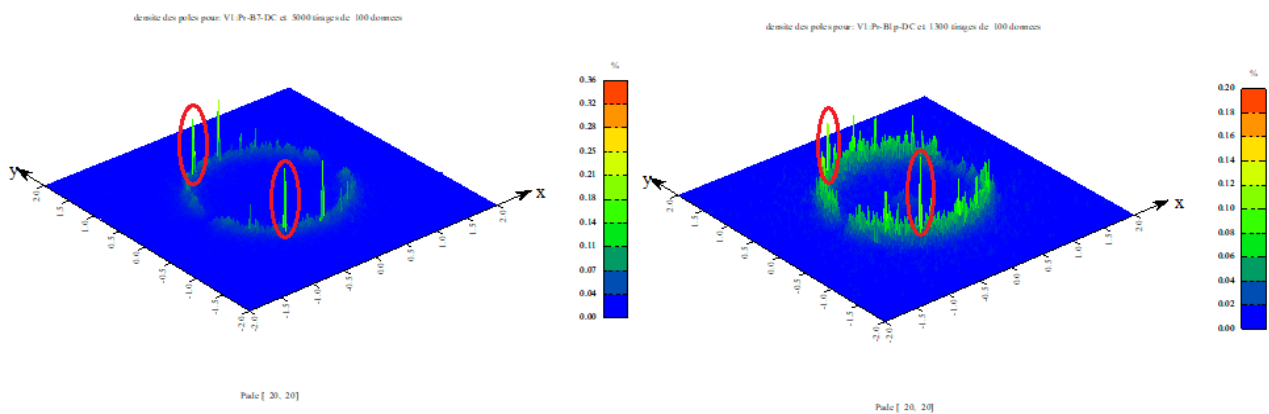

### 14. Prototypical analyticity patterns: T(z) and PA's to T(z).

Before proceeding with this description, we remind the reader that while our goal is to characterize the z-Transform in the complex, we rely in most concrete cases only on approximate hints pointed by PA's. Moreover while $T^{\mathcal{D}}(z)$ is a linear transform of the data, the PA is not linear and is admittedly continuous only in the capacity norm.

a. Constant signal
The data satisfy

$$d_n = d_0 \in \mathbb{R}^*$$

$$n = 0, \ldots N-1$$

and their z-Transform satisfies

$$T_N(z) = d_0 + d_0 z + \dots + d_0 z^{N-1} = d_0 \frac{1 - z^N}{1 - z} = \frac{1}{N} \frac{1 - z^N}{1 - z} T_N(1),$$

$$T_N(1) = N d_0.$$

The zeros are regularly placed on the unit circle. In the asymptotic regime $|z|^N \ll 1, |1 - z| \ll 1$, which is not empty if N is large enough, $T_N(z)$ becomes a [0/1] fraction with a simple pole located at 1 with residue ($-d_0$). This rational function is well represented by PA's. See Figures (14.a.i), (14.a.ii).

      b.   Sampled deterministic damped oscillatory signal.

Up to a phase, the data in this case are

$$d_n = d_0\, e^{-\ln 2\, n\tau/H} \cos\left(2\pi n \tfrac{\tau}{P}\right),$$
n = 0, 1, …, N − 1
$d_0 \in \mathbb{R}^*$,
$\tau$ sampling time,
P quasi period,
H half-life or decay time.

The z-Transform of the N-data set reads

$$T_N(z) = \frac{d_0}{2} \frac{1 - a^N z^N}{1 - az} + \frac{d_0}{2} \frac{1 - b^N z^N}{1 - bz}$$

where

$a = e^{-\ln 2\, \tau/H}\, e^{2i\pi\tau/P}$,
$b = e^{-\ln 2\, \tau/H}\, e^{-2i\pi\tau/P} = \bar{a}$.

In the asymptotic regime $\left| a^N z^N \right| = \left| b^N z^N \right| \ll 1$, $T_N(z)$ is well represented by

$$T_N(z) \simeq -\frac{d_0}{ab} \frac{\frac{a+b}{2} z - 1}{\left(z - \frac{1}{a}\right)\left(z - \frac{1}{b}\right)}$$

which has one real zero at

$$z_* = \frac{2}{a + \bar{a}} = \frac{e^{\ln 2\, \tau/H}}{\cos\left(2\pi \tfrac{\tau}{P}\right)}$$

and two simple poles at $(1/a)$ and $(1/\bar{a})$ with respective residues $(-d_0/2a)$ and $(-d_0/2\bar{a})$. See figures (14.b.i), (14.b.ii) for the PA representation.

Figure 14.a.i

Figure 14.a.ii

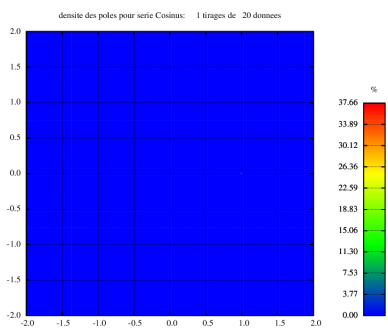

*Pole density for a Constant signal.*

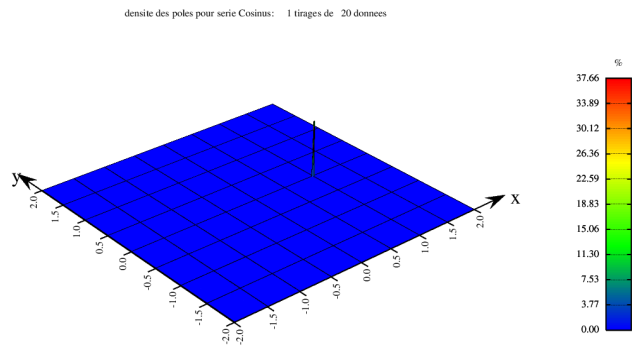

*Landscape of the pole density for a Constant signal*

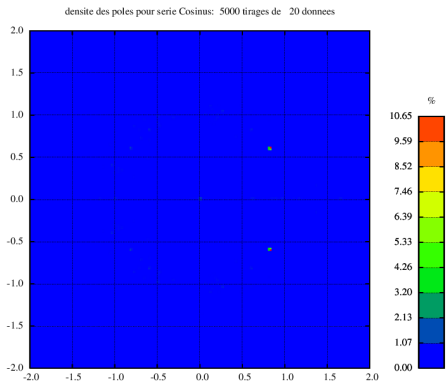

*Pole density for a Cosinus signal*

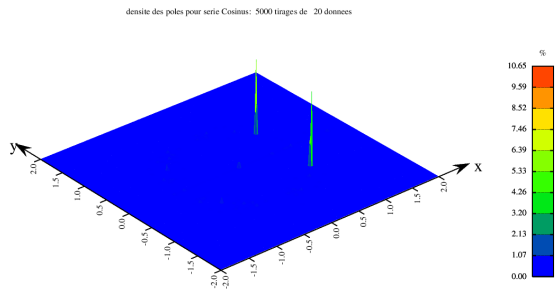

*Landscape of the pole density for a Cosinus signal*

c.   Random stationary gaussian white process.

The numerical Padé Filtering reveals a two components pole distribution supported by the real axis $\mathbb{R}$ and a fuzzy unit circle in $\mathbb{C}$. There are peaks at $-1$ and $+1$. On the circle the density is quite uniform except near $\pm 1$. See Figures 14. c.i and 14.c.ii.

Figure 14c.i

Figure 14.c.ii

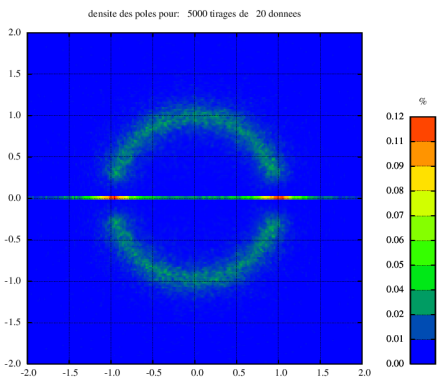

*Pole density for White Gaussian noise*

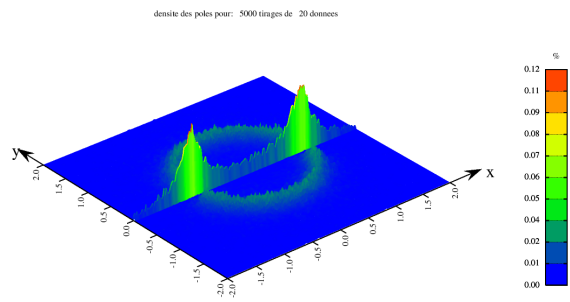

*Landscape of the pole density for White Gaussian noise*

d.   Padé Filtering of Random stationary colored gaussian processes: ARMA generated gaussian pink noise AR(1) and gaussian red noise AR(2).

The circular component of the pole density is no more uniform. A sizeable fraction of the poles is 'confined' by the resonances and the inverses of the resonances of the system. Peaks still exist at $\pm 1$ but they are modulated. See Figures 14.d.(i), (ii), (iii), (iv).

Figure 14.d.i

Figure 14.d.ii

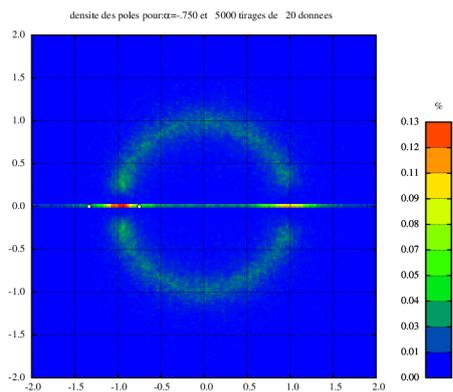

density des poles pour xr=-750 et   5000 tirages de   20 donnees

*Pole density for AR(1) series*

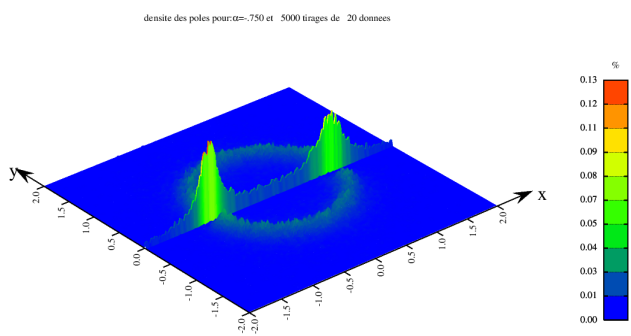

density des poles pour xr=-750 et   5000 tirages de   20 donnees

*Landscape of the pole density for AR(1) series*

Figure 14.d.iii

Figure 14.d.iv

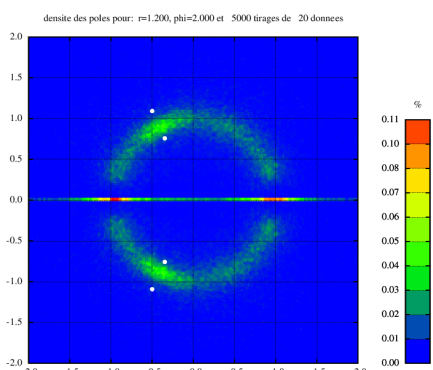

density des poles pour:   r=1.200, phi=2.000 et   5000 tirages de   20 donnees

*Pole density for AR(2) series*

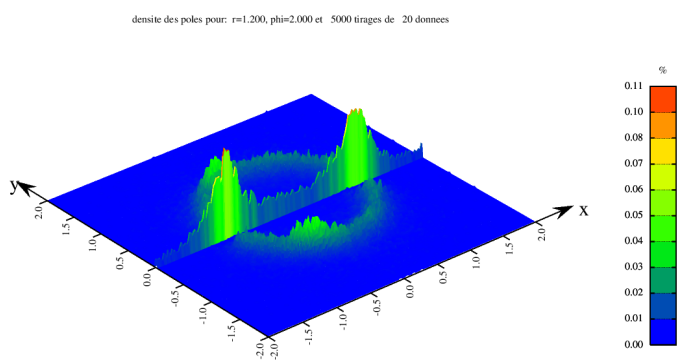

density des poles pour:   r=1.200, phi=2.000 et   5000 tirages de   20 donnees

*Landscape of the pole density for AR(2) series*

e.    Deterministic oscillatory signal with a random gaussian white background noise.

The features of the Padé Filtering picture depend on the relative amplitude of the signal and the noise. For equal or double amplitude of the cosine with respect to the noise, the characteristic peaks of the oscillation are present albeit thicker; the real and circular components which are characteristic of the stochastic part are also present. They almost disappear when the cosine is ten times larger than the noisy background. As noticed by the Houston group in all cases the uniformity of the circular distribution is broken. See Figure 14.e.(i), (ii), (iii), (iv), (v), (vi).

Figure 14.e.i

Figure 14.e.ii

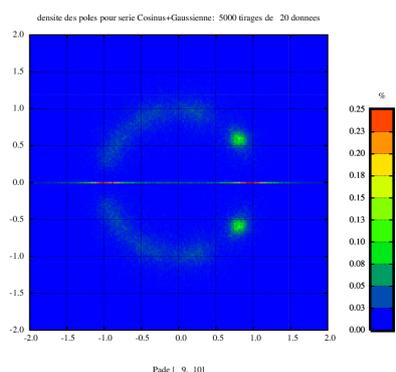

density des poles pour serie Cosinus+Gaussienne:   5000 tirages de   20 donnees

*Pole density for Cosinus sequence to Gaussian noise (with same amplitude)*

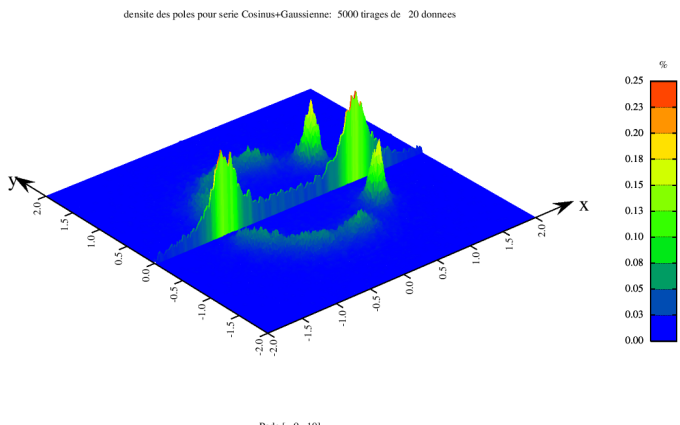

density des poles pour serie Cosinus+Gaussienne:   5000 tirages de   20 donnees

Pade [   9,  10]

*Landscape of the pole density for Cosinus sequence added to  added Gaussian noise (with same amplitude)*

Figure 14.e.iii

Figure 14.e.iv

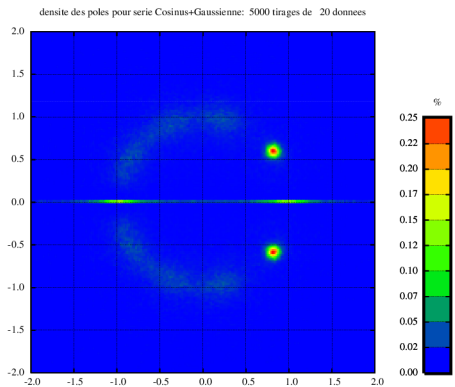

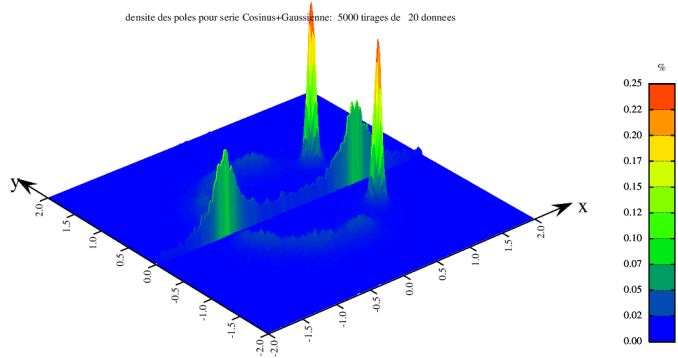

*Pole density for Cosinus sequence added to Gaussian noise (with cosinus 2 times the White Gaussian amplitude)*

*Landscape of the pole density for Cosinus sequence added to Gaussian noise (with cosinus 2 times the White Gaussian*

Figure 14.e.v

Figure 14.e.vi

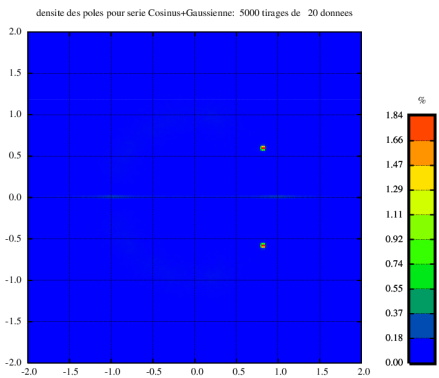

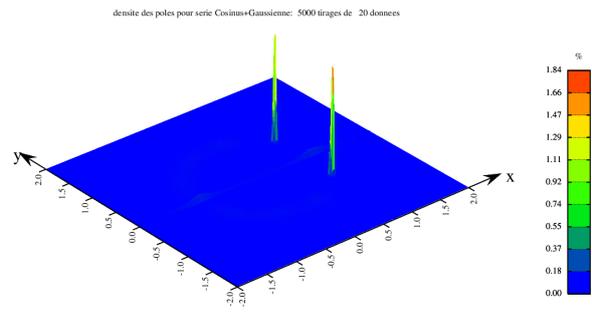

*Pole density for Cosinus sequence added to Gaussian noise (with cosinus 10 times the White Gaussian amplitude)*

*Landscape of the pole density for Cosinus sequence added to Gaussian noise (with cosinus 10 times the White Gaussian*

## VII. Padé filtering in the making

To investigate more closely the potential of the method, we test it in the setting of interferometric GW detection. Although we believe the method is more suitable for the assessment of noisy situations, we test it both on deterministic astrophysical signals and on noisy channels.

### 15. Testing on signals that are expected from the Gravitational Universe.

We concentrate on the theoretical models of the GW's produced by coalescing binaries (CB) and by rotating magnetized neutron stars (pulsar).

<u>CB</u> The two elements of the binary system may be a black hole (BH) or a neutron star (NS). The two may orbit around each other for a long time, with their relative distance slowly decreasing and with negligible GW emission; nonetheless, when their distance reaches the order of magnitude of their radius, the process becomes catastrophic and a burst of GW's is emitted; this burst exhibits three different phases: late inspiral chirp signal, merging and ringdown. In the chirp phase, the amplitude and frequency of the GW signal both increase to potentially become infinite; this process stops when the two bodies actually merge, to be replaced by a relaxation process of the new body; continuity of the signal and of its time derivative at the merging instant ensure the link between the first and last phases. The corresponding templates are amenable to General Relativity perturbative (possibly renormalized) calculations. Here we limit ourselves to lowest order models (Th. Damour 2016) to which we apply Padé Filtering, namely

| chirp | $h(t) = c_1(t_0 - t)^{-\frac{1}{4}} \cos\left[c_2(t_0 - t)^{\frac{5}{8}} + c_3\right]$ | $t \leq t_c$ |
|---|---|---|
| merging | | $t = t_c$ with $t_c < t_0$ |
| ringdown | $h(t) = A \exp\left[-\pi \dfrac{F}{Q} \lvert t - t_0 \rvert\right] \cos[\varphi_0 - 2\pi F(t - t_0)]$ | $t_c \leq t$ |

Concerning the chirp phase see the numerical results Figures 15.1, 15.2 and 15.3.

Figure 15.1

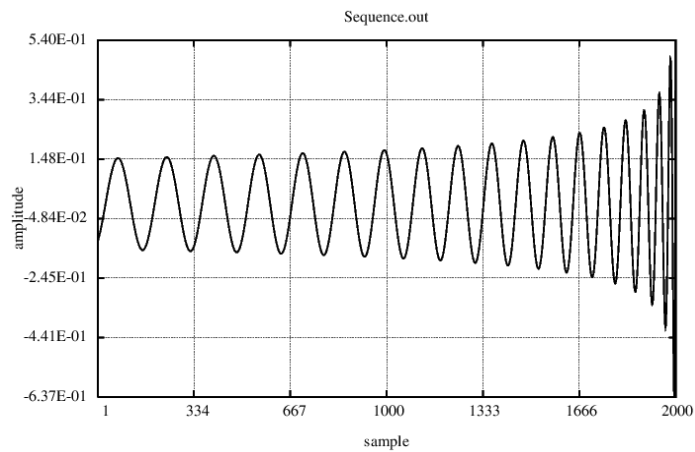

Figure 15.2

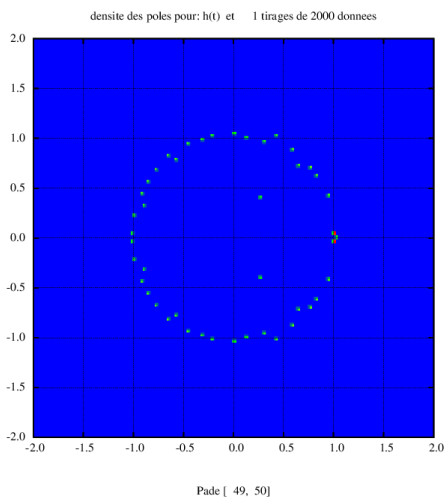

Figure 15.3

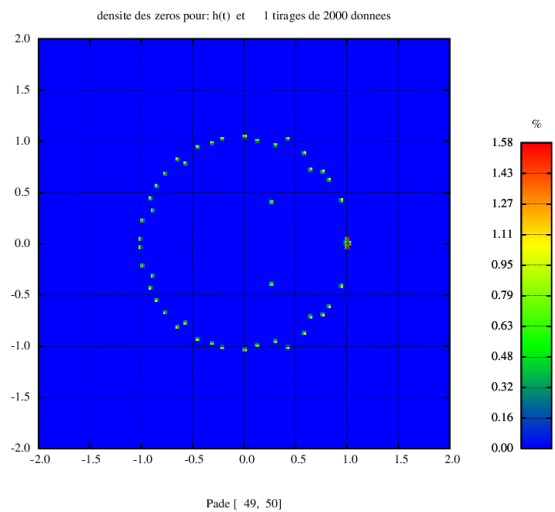

The ringdown phase can be worked out. As an example we consider the sub-phase $t_0 \leq t$ in the large $N$ régime, where $N$ is the number of data; we denote $\tau = \frac{1}{F_s}$ the time discretization step; $T_N(z)$ is then approximated by a $[1/2]$ fraction with one zero $z_0$ and a pair of complex conjugate poles $z_p$ and $\overline{z_p}$ , with

$$z_p = \exp\left[\pi \frac{F}{F_s Q}\right]\exp\left[2i\pi \frac{F}{F_s}\right],$$

$$z_0 = \exp\left[\pi \frac{F}{F_s Q}\right]\frac{\cos(\varphi_0)}{\cos\left(2\pi\frac{F}{F_s}+ \varphi_0\right)}.$$

The PA $[L/M]$ to this fraction is the fraction itself; figure 15.4 and 15.5 illustrate these findings.

Figure 15.4

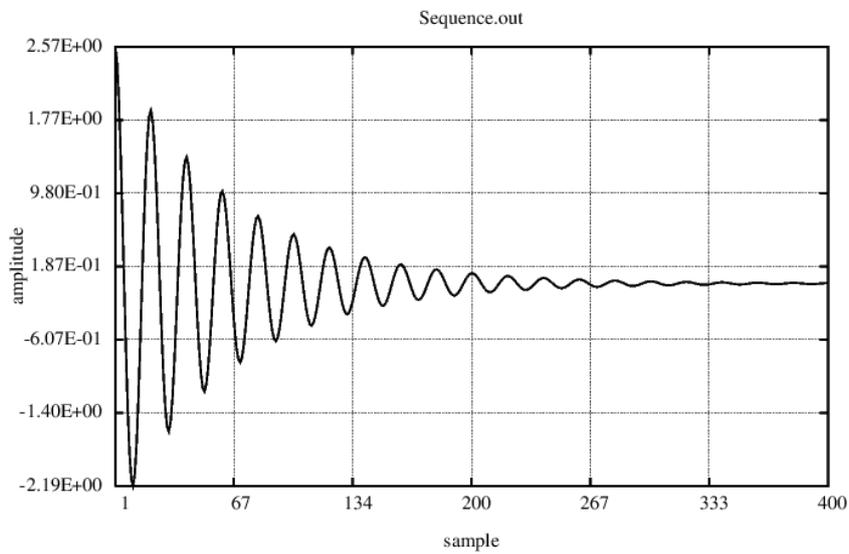

Figure 15.5

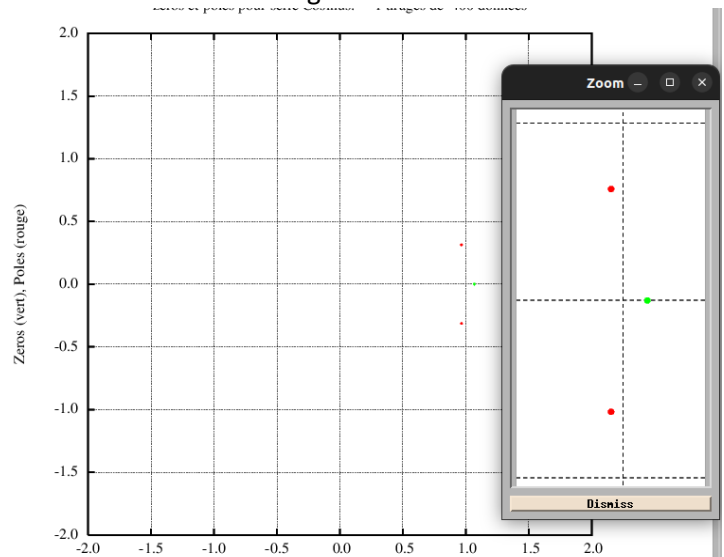

<u>Pulsar</u> While the coalescence of two compact objects produces a GW blow-up, a rotating magnetized neutron star emits a feeble but continuous GW signal. This emission causes the rotational kinetic energy to decrease without changing the internal structure of the body, thus leading to a slowing down of the rotation and therefore of the GW frequency $F$. Assuming that the star's rotation axis is perpendicular to the line of sight, and considering a given GW polarization, the detected signal has the form

$$h(t) \propto \cos \Phi(t)$$

where $\Phi(t)$ can be evaluated by a Taylor expansion. To the first not trivial order it can be written

$$\Phi(t) = 2\pi F t (1 - \frac{1}{m} 10^{-b} t)$$

where phase factors and Doppler effects have been dropped; assuming a rigid rotation of the neutron star and an energy loss due to GW emission only yields

$$m = 8,$$
$$b = 9.$$

Numerical tests show that $\mathcal{O}\left(10^{10}\right)$ is too large a value for the ratio of the two characteristic times of the model to be resolved by Padé Analysis. To have a first glimpse at the phenomenon in the complex, we resort to a model of the model in setting $b = 4$. After time sampling at a frequency $F_S$, the data read

$$d(j) = \cos\left[2\pi \frac{F}{F_S} j \left(1 - \frac{1}{8} 10^{-4} \frac{j}{F_S}\right)\right].$$

One then proceeds with the Padé Filtering; the distributions of poles and zeros shown in Fig. 15.6 were obtained for $F = 20$ Hz, $F/F_S = 1/10$, and the [50/50] Padé. We comment that (i) the main peak is accompanied by secondary peaks on the unit circle where zero's and pole's peaks are close ($\mathcal{O}\left(10^{-4}\right)$) to each other but distinct; (ii) the peak at the origin of the pole's distribution is spurious in the sense that the peak at the origin of the zero's distribution is extremely close ($\mathcal{O}\left(10^{-10}\right)$).

Figure 15.6. i

Figure 15.6.ii

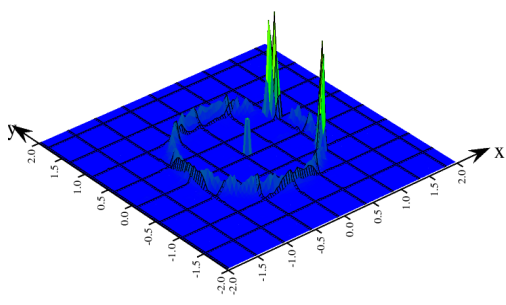
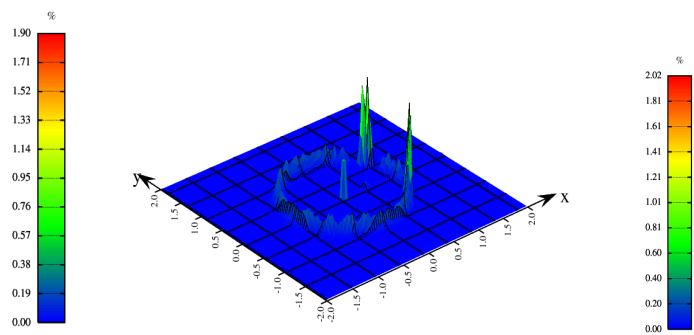

16.  Testing on Virgo output channels subject to environmental noises.

We now turn to the study of actual data collected as the Virgo interferometer[***]. We concentrate on the coupling between the state of the detector and the environmental perturbation it is subjected to. Among them the most critical one is the micro-seismicity which governs the bad low frequency response of the interferometer to incoming GW's; the end mirrors, for instance, are put in motion while they should be at rest and be 'moved' only by the extremely tiny GW's. This phenomenon has long been expected and specially designed hierarchical suspensions greatly damp this unwanted motion. We devote this section to the discussion of the link between the vertical component of the seismicity and the vertical components of the motion of the first and last stages of the suspension of the Power Recycling mirror. Technically these three components are monitored by the outer channel V1:ENV_CEB_SEIS_V_rms (432,000 data), and the V1:Sc_PR_F7_Z_rms and V1:Sc_PR_F0_Z_rms output channels (432,000 data each); we use various correlation functions and Padé Filtering techniques to characterize the relationship between these data.

---

[***] In the Bibliography the interested reader can find references on the interferometric detection of GW's.

Figure 16.1(i)

Figure 16.1(ii)

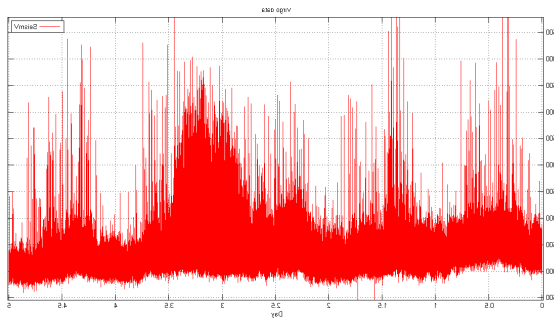
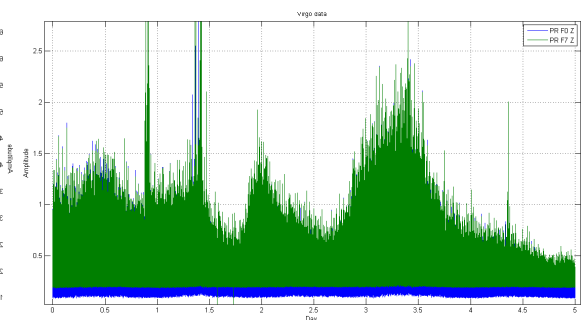

Looking at Figure 16.1(i and ii), which represents the data of the above cited channels for 5 consecutive working days, one immediately notices the gross 1-day periodicity; anthropic disturbances appear to be the common cause of all three motions. Apart from their very different order of magnitude, the signals look akin and the energy spectra of the second and third (see Fig. 16.2(i)) exhibit the same bumps which are likely to be the marks of resonances of the suspension; apart from the peak at the origin the spectrum of the seismic vertical motion (Fig. 16.2(ii)) does not indeed exhibit clean bumps.

Figure 16.2(i)

Figure 16.2(ii)

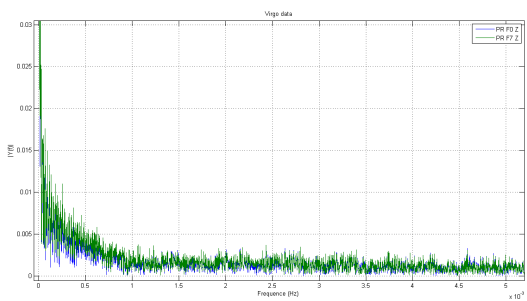
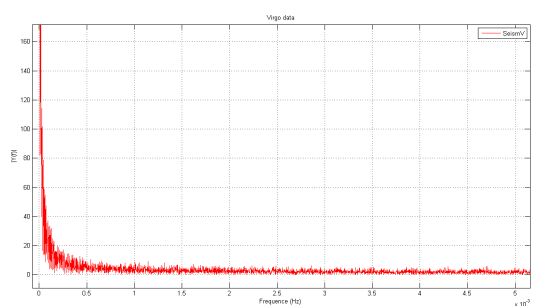

To further examine the links between these three signals we use their auto-and cross-correlation functions.

Fig. 16.3(i) and Fig. 16.3(ii) tell us about the close similarity between the motion of the first and last stages of the suspension in the small scale but nothing definite about their link with seismicity. These pictures also exclude a simple $\delta$-time-delay model between seismicity and the motions of the suspension; such a model would indeed imply a simple $\delta$-time-shift relation of the type

$$C_{1,2}(t) = C_{1,1}(t - \delta),$$

something which is ruled out by the shape of the correlation curves; this points towards a linear but more refined relationship between the signals; we now explore this relationship within the z-Transform formalism.

Figure 16.3(i)                                             Figure 16.3(ii)

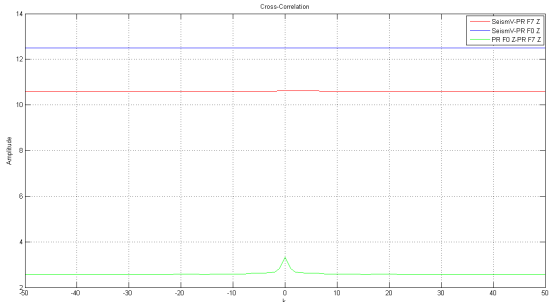    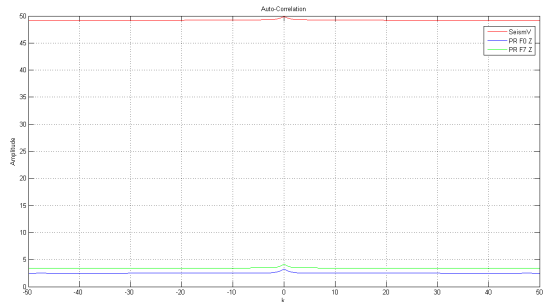

Figure 16.4(i)                                             Figure 16.4(ii)

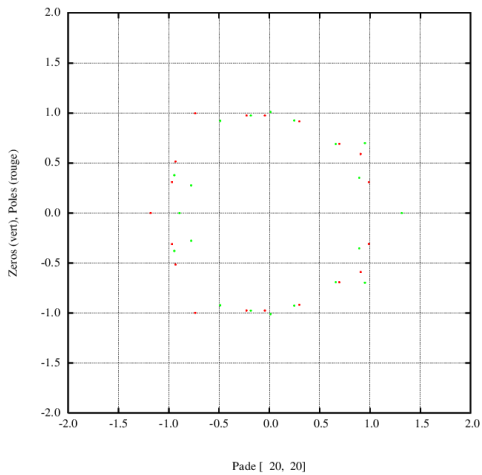    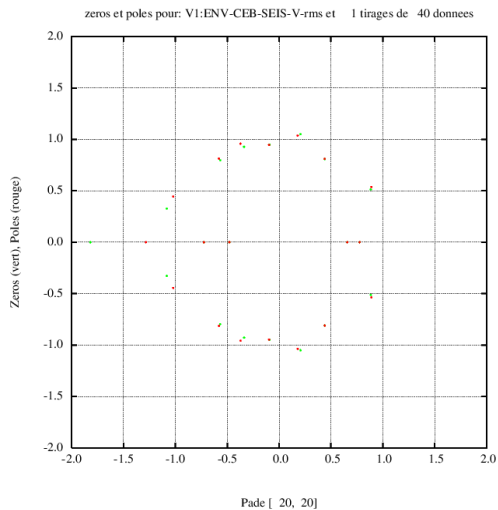

In doing so we want to clarify the way Padé Filtering distinguishes between two sets of data which are akin and are linearly related. We apply PF to the seismicity channel and the F0 channel; as usual we express the result in plotting the distribution on $\mathbb{C}$ of poles and zeros of the PA's of the z-Transforms of the data (Fig. 16.4 (i) and Fig. 16.4(ii)). Such pictures may be considered as images or as probability distribution functions on the plane. Adopting the first point of view we calculate the correlation factor between the two pole distributions; adopting the second point of view we calculate the Kullback-Leibler divergence of the V1:Sc_PR_F0_Z_rms, respectively V1:Sc_PR_F7_Z_rms, pole distribution from the seismic one; the results are presented in Table 16; they coherently and quantitively report on the situation previously described qualitatively.

Table 16

|      | Seismo-seismo | Seismo-F7 | F0-F0 | F0-F7 |
|------|---------------|-----------|-------|-------|
| Corr | 1             | 0.95      | 1     | 0.97  |
| KL   | $5.\,10^{-6}$ | 6.14      | $6.\,10^{-6}$ | 5.4 |

Finally we express the linear relation between the seismic channel and the V1:Sc_PR_F0_Z_rms channel in terms of the product of z-Transforms,

$$T_{F0} = T_K T_S \;;$$

$T_{F0}$ is the z-Transform of the output data, $T_S$ is the z-Transform of the seismic data and $T_K$ is the z-Transform of the stationary linear operator $K$ representing the action of the suspension on the seismicity. Approximating $T_{F0}$ and $T_S$ by their respective PA's

$$T_{F0} \simeq \frac{L_0}{M_0} ,$$

$$T_S \simeq \frac{L_S}{M_S},$$

the above equation yields

$$T_K \simeq \frac{L_0 M_S}{L_S M_0}.$$

We specify all these quantities in using e.g. PA's of the type [20/20] with one realization. The approximated $T_K$ looks like having 40 zeros and 40 poles, whose locations are depicted in Fig. 16.4(i) and Fig. 16.4(ii).
But as a matter of fact a lot of those locations are quasi-common to the two pictures and thus many zeros and poles compensate each other or enter in doublets; we 'erase' those doublets to keep only the 'skeleton' of $T_K$ which turns to be a [5/5] fraction; up to a check on other chunks of data, this performs an information compression concerning the way the interferometer suspension reacts to the environmental seismic noise (Fig. 16.5).

Figure 16.5

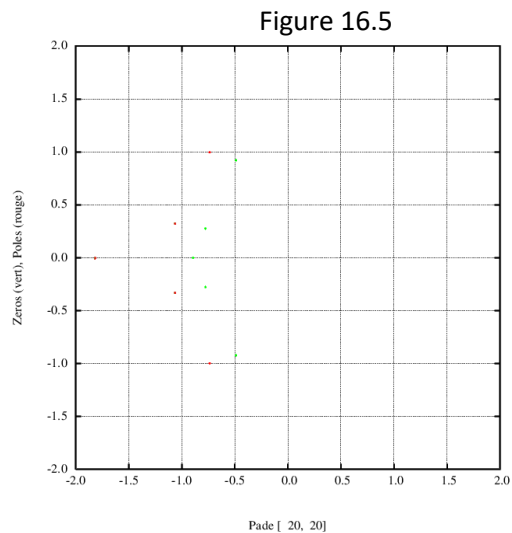

Pade [ 20, 20]

17. Actual implementation. Generalizations. The Froissart debate.

Actual implementation.

As for any processing of actual data collected at an interferometric GW detector, we advise to perform the routine preprocessing, namely:
- undersampling from the usual 20 kHz to 1 kHz
- bandpass filtering [20 Hz, 500 Hz]

- remove the 50 Hz component (60 Hz in the US), the 24 hours periodicity, as well as the known noise components due to seismicity or mechanical resonances
- try to understand as much as possible the resulting signal using classical tools like Fourier Transform and correlation functions
- etc…

Then the PF itself should be performed with special care, namely:
- check for the existence of freak zeros and poles, located outside the 2D histogram window
- remove all the real poles, which especially ensures a zero mean signal
- compare the location of the poles and zeros in order to sort the doublets into extension magnitude; the doublets may be completely spurious, or reflect the various noises
- try to understand the unexpected Padé Table
- etc…

One is then in position to extract PF based informations, namely:
- calculation and visualization of the residues of the poles
- recognition of analyticity patterns
- state of the detector
- presence of a certain noise and possibility to erase it
- cross-correlation and distance between distinct channels singularity structure
- etc…

Generalization.

The complete localization of a source in the sky needs a joint detection by at least two antennas. On the other hand, one may wish to determine the local or global character of -anthropic or natural- disturbances suspected at separate antennas. To be specific let us assume we have two separate antennas both capable to detect a certain transient ringdown. One considers the double data collection model

$$s_1(t) = n_1(t) + A_1 e^{-\alpha t} \cos(\omega t + \varphi_1),$$

$$s_2(t) = n_2(t) + A_2 e^{-\alpha t} \cos(\omega t + \varphi_2),$$

where $n_1(t)$ and $n_2(t)$ are independent noises. Following the main stream PF line, one would try to jointly study the z-transforms of $s_1$ and $s_2$, each according to §15. In a Phys. Rev.D (2014) paper, L. Perotti, T. Regimbau, D. Vrinceanu and D. Bessis discussed the new idea of complexifying the data[†††] according to

$$s(t) = s_1(t) + i s_2(t)$$

and studying the z-transform of $s(t)$. Since one has to calculate one family of PA's, instead of two, this method is computationaly less costly. The locations of the poles in $\mathbb{C}$ is more involved, but the authors show that in concentrating on the lines of minimum distance between quasi complex conjugate poles, the detection of the coincident GW transient is feasible.

The Froissart debate.

The clear separation between the faithful poles and zeros and the spurious ones, forming the Froissart doublets, is the corner stone of the Padé Filtering method. But the very existence of those doublets is a radical obstacle to any pointwise large order convergence of PA's; so much so that mathematicians have tried to get rid of them at the cost of changing the very definition of PA's, introducing 'robust Padé Approximants' via Singular Value Decomposition. But even so, the situation is not clear cut and is still under discussion.

---

[†††] We remind the reader that until now we considered only real data.

# COMPLEMENTS AND APPENDICES

# COMPLEMENTS AND APPENDICES

## IX. Reminder on classical Padé Approximants theory

To revisit the topic of complex singularities, we suggest the course "Dealing with the singularities of analytic functions" by J.-D. Fournier and D. Bessis; this course is a section of the proceedings of a mathematical physics winter school (Chamonix 1993, see Bibliography); it includes a short simple chapter on Padé Approximants. For some of the proofs concerning PA's we suggest the lecture given by E. Saff at the Porquerolles summer school (2005), see bibliography, the book by G. A. Baker Jr (from which we borrow the notations used here) and the book by J. Gilewicz.

The first idea is to devise a better approximant to a given function than the truncated Taylor expansion of this function. The goal may be to improve the precision of the evaluation of the function or the description of its analytic structure; we concentrate here on the latter. Being a finite degree polynomial, the Taylor approximant is indeed always singular at infinity and never singular at a finite distance. To circumvent this pitfall one resorts to a fractional approximant and the approximation criterium is written in terms of the Taylor expansions of both the rational approximant and the approximated function. Precisely, let us consider two polynomials

$$P_L(z) = p_0 + p_1 z + \cdots + p_L z^L$$

$$Q_M(z) = 1 + q_1 z + \cdots + q_M z^M$$

of degree at most $L, resp. M$, and without common zeros. The ratio

$$[L/M](z) = \frac{P_L(z)}{Q_M(z)}$$

is called the $(L/M)$ Padé Approximant to the function $f(z)$ if the condition

$$Q_M(z)f(z) - P_L(z) = \mathcal{O}(z^{L+M+1})$$

is fulfilled. Practically, $f(z)$ is represented by its (possibly formal) Taylor expansion known up to a degree $N$

$$f(z) = \sum_{0 \le j \le N} d_j z^j + \mathcal{O}(z^{N+1}).$$

So for given $N$, the definition makes sense if

$$L + M \le N.$$

The $d_j's$ being known, the PA condition gives rise to a set of linear equations for the $p$'s and the $q$'s

$$
\begin{aligned}
d_0 &= p_0 \\
d_1 + d_0 q_1 &= p_1 \\
\dots &\quad \dots \\
d_L + d_{L-1} q_1 + \ldots + d_0 q_L &= p_L \\
d_{L+1} + d_L q_1 + \ldots + d_{L-M+1} q_M &= 0 \\
\dots &\quad \dots \\
d_{L+M} + d_{L+M-1} q_1 + \ldots + d_L q_M &= 0
\end{aligned}
$$

If a solution exists, it is unique. The resulting $(L/M)$ PA can be explicitly written in terms of determinants but the numerical calculation of them is unstable. This is one of the motivations for mentioning a bit of the properties of the so called Padé Table.

The various $[L/M](z)$ PA's to a given function $f(z)$ can be arranged according to the following table

| [0/0] | [0/1] | ... |
|-------|-------|-----|
| [1/0] | [1/1] | ... |
| ... | ... | $[L/M]$ |

called the Padé Table. Here we leave aside the important discussion of its possible singularities, which occur when successive entries of the table are identical; those entries then form square blocks, which require a proper treatment when exploring the Padé Table. If the Padé Table does not have blocks, it has an internal structure expressed by invariants and recursion relation among the PA's. We quote two of them, the Wynn's identity

$$\frac{1}{[L+1/M]-[L/M]} + \frac{1}{[L-1/M]-[L/M]} = \frac{1}{[L/M+1]-[L/M]} + \frac{1}{[L/M-1]-[L/M]}$$

and the recursion relation which allows the calculation of $[L/M]$ knowing its neighbors $[L/M-1]$ and
$[L+1/M-1]$. Iteration of such transforms allows to calculate recursively all the members of a given para-anti-diagonal of the Table, hence the whole Table.

In the present document we are mainly interested to the location of zeros and poles of the PA's. As explained in §4, §10 and §11, "defects" or "doublets" occur in their repartition, in particular if $f(z)$ is random. This is the cause for involved convergence properties in the limit $L$ and/or $M$ large as developed by Baker quoted hereafter:

*The discussion of the convergence of Padé approximants is complicated by the occurrence of "defects". By this is meant a zero and a pole which are very close to each other. The value of the Padé approximant is disrupted over a very small region. Much of the effort on the convergence of Padé approximants relates to whether this problem occurs in a specific case and if it does how to limit the size of the disruption. Thus there is pointwise convergence, convergence in measure, convergence in Hausdorff measure, convergence in capacity, and convergence in the mean. In many cases, the diagonal Padé approximants are the most effective sequences.*

# X. Reminder on basic probability calculations

We consider here a real random variable $Z$ (rrv) such that

$$\mathcal{P}_Z(x, dx) \equiv probability\ \{x \le Z \le x + dx\} = d\mu_z(x) = p_z(x)dx$$

where $\mu_z(x)$ is a positive normalized measure having a density $p_z(x)$ which is itself a function or a Schwartz distribution. As an exercise we establish the probabilistic properties of the sum $W$ of two independent random variables $U$ and $V$

$$\mathcal{P}_W(w, dw) = Sum\ [\mathcal{P}_U(u, du)\mathcal{P}_V(v, dv)]$$

with the constraint $W = U + V$.

The rhs of the constrained equality can be written

$$Sum\ [\mathcal{P}_U(u, du)\mathcal{P}_V(w - u, dv)] = \int dw\ du\ p_U(u)\ p_V(w - u)\,;$$

the density of W thus reads

$$p_W(w) = \int du\ p_U(u)\ p_V(w - u).$$

Similarly one establishes for the product $W$ of $U$ by a fixed real not vanishing scalar $\lambda$

$$p_W(w) = \frac{1}{|\lambda|} p_U\left(\frac{w}{\lambda}\right).$$

<u>Notation for the first centered moments</u>

$m \equiv \int dx\ x\ p_X(x)$ is the mean of $X$

$\sigma^2 \equiv \int dx\ (x - m)^2\ p_X(x)$ is the variance of $X$.

<u>Gaussian variable</u>

To model a random phenomenon which is concentrated near its mean value it is customary to assume that its probability density has the profile of the Gauss bell function. The positivity, the normalizability and the fast decreasing, for large excursion, of this function make it a possible candidate; for natural phenomena it turns out to be to a large extent <u>the</u> candidate. This is due to the following circumstances:

(i) using the above definition and the normalization to 1 of the bell profile, the density $p_X(x)$ reads

$$p_X(x) = \frac{1}{\sigma} \frac{1}{\sqrt{2\pi}}\ exp\ \left[-\frac{1}{2} \frac{(x - m)^2}{\sigma^2}\right],$$

$$m \in \mathbb{R}\,,\ \sigma \in \mathbb{R}_+^*,$$

and $X$ is said to be gaussian;

(ii) the class of Gauss function is invariant under convolution; precisely

$$exp\left[-\frac{1}{2}x^2\right] \otimes exp\left[-\frac{1}{2}x^2\right] = \sqrt{\pi}\, exp\left[-\frac{1}{4}x^2\right];$$

(iii) the class of Gauss functions is obviously invariant under multiplication of the rrv $X$ by a scalar
$\lambda \in \mathbb{R}^*$;

(iv) this implies that a sum of two independent gaussian variables is itself a gaussian variable; so also a
gaussian variable multiplied by a scalar; so also any linear combination of independent gaussian variables;

(v) central limit theorem (CLT). Consider the sum $S_N$ of $N$ independent rrv with the same density $p_X(x)$, not necessarily a gaussian function; if the zero, first and second moments of $p_X(x)$ do exist, the rrv $S_N$ becomes gaussian in the large $N$ limit;

(vi) a meaning can be given to a "gaussian vector", made of $k$ rrv $X_j$, each being separately gaussian,
but possibly dependent. Multiplying a gaussian vector by a deterministic matrix produces another gaussian vector;

(vii) and more.

## XI. Bibliography

**Classical Complex Analysis and Applications**

<u>Books</u>

Functions of a Complex Variable : Theory and Technique
G.F. Carrier, M. Krook, C. E. Pearson (1966)

The Taylor Series
P. Dienes (1957)

<u>Section of a Book</u>

Dealing with the Singularities of Analytic Functions
J.-D. Fournier and D. Bessis
In An Introdution to methods of Complex Analysis and Geometry for Classical Mechanics
and Nonlinear Waves
Ed. D. Benest, Publ. Editions Frontières (1993)

**Classical Probabily Theory and Applications**

<u>Books</u>

An Introduction to Probability Theory and its Applications
W. Feller (1968)

Probability Theory : an Introductory Course
Y.G. Sinaï (1992)

Probability and Related Topics in Physical Sciences
M. Kac (1957)

Some Random series of Function
J.P. Kahn (1985)

On Lévy laws see chapt. 1 and Appendix B in Anomalous diffusion in Disordered Media
J.-Ph. Bouchaud and A. Georges, Physics Reports <u>195</u>, 4-5 (1990)

**Kac Phenomenon**

<u>Research Paper</u>

On the Average Number of Real Roots of a Random Algebraic Equation
M. Kac, Bull. Amer. Math. Soc. <u>49</u>, (1943), p. 314 with correction <u>49</u>, (1943) p. 938

**Froissart Phenomenon**

<u>Research Paper *in French*</u>

Approximation de Padé : Application à la Physique des Particules elémentaires
M. Froissart
in Les Rencontres Physiciens-Mathématiciens de Strasbourg
RCP25 (1969) <u>9</u>, 2, p. 1-13 *out of print*
*Available on internet*
http://www.numdam.org/article/RCP25_1969__9__A2_0.pdf

**Padé Approximants, Orhogonal Polynomials, Resonances in Physics**

<u>Books</u>

> Essential of Padé Approximants
> G.A. Baker Jr. (1975)
>
> Approximants de Padé *in French*
> J. Gilewicz (1978)
>
> Advanced Mathematical Methods for Scientists and Engineers
> C. M. Bender and S. A. Orszag (1978)
>
> Harmonic Analysis and Rational Approximation: Their Roles in Signals, Control and Dynamical Systems (Proceedings of the Porquerolles School 2003)
> Ed: J.-D. Fournier, J. Grimm, J. Leblond, J.R. Partington (2006)
>
> Orthogonal Polynomials
> G. Szegö (1939)
>
> Random Polynomials
> A.T. Bharucha Reid and M. Sambandhan (1986)

<u>Pdf Internet Document</u>

> Introduction to Padé Approximants
> E. Saff Lecture at the Porquerolles School (2003)
> http://www-sop.inria.fr/apics/anap03/padeTalk.pdf

**Polynomials and Fractions in Data Analysis**

<u>Books</u>

> Modern Spectal Estimation: Theory and Application
> S. M. Kay (1987)
>
> Time Series Analysis : Forecasting and Control
> E.P. Box, G.M. Jenkins, G. C. Reinsel, G. H. Ljung (2015)
>
> Extrapolation, Interpolation and Smoothing of Stationary Time Series
> N. Wiener (1949)

<u>Section of a book</u>

> On the Role of Orthogonal Polynomials on the Unit Circle in Digital Signal Processing Applications
> P. Delsarte and Y. Genin in Orthogonal Polynomials: Theory and Practice, Ed. P. Nevai (1990)

**Gravitational Waves and their Interferometric Detection**

<u>Book</u>

> Gravitation
> C. W. Misner, K. S. Thorne, J. A. Wheeler (2017)

<u>Review Papers</u>

> Gravitational Waves and Binary Black Holes
> Th. Damour, Séminaire Poincaré XXII (2016)

An Aperçu about Gravitational Waves and Data Analysis
E. Chassande-Mottin, Séminaire Poincaré XXII (2016)

<u>Research Paper</u>

Observation of Gravitational Waves from a Binary Black Hole Merger
B. Abbott et al.
Phys. Rev. Letter, <u>116</u>, 061102 (2016)

**Froissart Debate**

Robust Padé Approximation via SVD
P. Gonnet, S. Guttel, L. N. Trefethen
SIAM Review, <u>55</u> (2013) 1 p. 101-117

Algebraic Properties of Robust Padé Approximants
B. Beckermann and A. Matos
J. App. Th. <u>190</u> (2015) p. 91-115

Robust Padé Approximants May Have Spurious Poles
W. F. Mascarenhas
J. App. Th. <u>189</u> (2015) p. 76-80

On Rational Functions Without Froissart Doublets
D. Beckermann, G. Labahn, A. C. Matos
Num. Mathematik <u>138</u> (2018) 3 p. 615-633

**Research Articles alluded to in the document**

Asymptotics of Zeros of Orthogonal and Paraorthogonal Szegö Polynomials in Frequency Analysis
W. B. Jones, O. Njåstad and H. Waadeland
In Continued Fractions and Orthogonal Functions
Ed. S. C. Cooper and W. J. Thron, Publ. Taylor and Francis Group (1993)

Padé Approximations in Noise Filtering
D. Bessis
JCAM <u>66</u> (1996) p. 85-88

Padé Approximants and Noise: A Case of Geometric Series
J. Gilewicz and M. Pindor
JCAM <u>87</u> (1997) p. 199-214

Padé Approximants and Noise: Rational Functions
J. Gilewicz and M. Pindor
JCAM <u>105</u> (1999) p. 285-297

On the Distribution of Poles of Padé Approximants to the z-Transform of Complex Gaussian White Noise
P. Barone
J. App. Th. <u>132</u> (2005) p. 224-240

The Footprints of Noise in Rational Interpolation
J.-D. Fournier, M. Pindor
Int. J. of Modern Phys. D <u>09</u> (2000) 03 p. 281-285

Rational Interpolation from Stochastic Data: A New Froissart Phenomenon
J.-D. Fournier, M. Pindor
Reliable Computing <u>6</u> (2000) 4 p. 391-409

# XII. Captions of the Figures

Figure 3.1    Shape of the Lorentz single line spectrum for various half widths $\Gamma$

Figure 3.2    Shape of a bi-Lorentz spectrum as the common half width $\Gamma$ of the two lines varies while their separation is kept constant. For small and moderate $\Gamma$ there are two maxima; for $\Gamma$ large enough the two maxima are smoothed out and the two lines are no longer distinguishable. At the same time the singularity complex structure made of four poles is kept invariant.

Figure 4.1    Numerical illustration of the Kac phenomenon for finite real polynomials: comparison between the experimental histogram of the real root's density $\rho_n(t)$ and Kac' prediction. The density is peaked near $\pm 1$. The polynomial is a homogeneous random polynomial of degree $n = 10$.

Figure 4.2    Same as Figure 4.1 with a polynomial of degree $n = 100$

Figure 4.3    Numerical illustration of the Kac phenomenon for finite real polynomials of various degrees. The average number of real roots is plotted against the degree $n$ of the random polynomial, both experimentally (dots) and theoretically (lines). The diagram is in a $log - lin$ scale to visualize the asymptotic logarithmic behavior. Whatever the degree a finite fraction of the root is real.

Figure 5    Auxiliary calculation. Plot of the transition curve defined by $\rho = 1 - |sin\theta|$ in polar coordinates. The asymptotic ($\rho \lesssim 1$) behavior of the gaussian determinant $\Delta$ breaks down when the point ($\rho, \theta$) approaches this curve.

Figure 7.1    Statistical distribution of the roots of the random polynomial $P_n(z) = \Lambda z^n + \sum_{k=0}^{n-1} a_k z^k$ where $n = 10$, $\Lambda = 20$, the $a_k$ are rrv iid $\mathcal{N}(0,1)$ and $\mathcal{M} = 50000$ draws have been performed. The roots are fluctuating in the complex near the 20 nodes of a circular quasicrystal.

Figure 7.2    As Figure 7.1. The logarithm of the radius of the circular quasi crystal is plotted against $(1/n)$, with $\Lambda = 20$.

Figure 7.3    A Figure 7.1 but $\Lambda = 0$, $n = 10$, $\mathcal{M} = 50000$. There is no quasi crystal in this Kac case.

Figure 8.1    Mean energy spectrum $\mathbb{E}[\mathcal{E}_\infty^X(\theta)]$ of the AR(1) pink noise model as a function of $\theta$ and for various values of the parameter $a$; the length $N$ is infinite. Whatever the value of the parameter $a$, the peak is located at the origin. The z-Transform has a moving outer pole in $(1/a)$. Red $a = 0.3$. Blue $a = 0.6$. Green $a = 0.8$.

Figure 8.2    Auxiliary calculation. Plot in the parameter $(\varphi, r)$ plane of the transition curve $r_+(\varphi)$. Above the curve the mean energy spectrum of the AR(2) red noise model has one peak at the origin ($\theta = 0$); below it has two peaks at $\pm\theta_*$.

Figure 8.3    Mean energy spectrum of the AR(2) noise as a function of $\theta$ for various positions of the parameter in the $(\varphi, r)$ plane. For $\varphi = 0$, there is always only one peak located at the origin. For $\varphi = \frac{\pi}{2}$ there are always two peaks located at $\pm\frac{\pi}{2}$. For $\varphi$ in between, the situation depends on the position of the parameter $(\varphi, r)$ with respect to the transition curve $r_+(\varphi)$ depicted in Figure 8.2. The z-Transform has two moving outer poles in $r \exp(\pm i\varphi)$.

Figure 10    Artist's view of the discovery of the Froissart doublets (FD) in the RI to $\varphi(z) = \tanh z$. The faithful zeros and poles are correctly spaced on the imaginary axis. The spurious ones, forming the FD's, are located inside or near the interpolation interval.

Figure 13    Proof of concept of Padé Filtering. The peaks of the polar distribution of the channel *end of north arm* reappear in the polar distribution of the channel *dark port*. Padé $[20/20]$, $\mathcal{M} = 1300$ draws.

Figures 14.a.i and 14.a.ii
Padé Filtering of a constant signal of length $N = 20$. The density of pole of the z-Transform of the signal is plotted as a function of $z \in \mathbb{C}$. The poles are obtained via $[10/10]$ Padé Approximants. The plot exhibits one peak at 1. The figure (ii) is a 3D perspective or "landscape" view of this peak.

Figures 14.b.i and 14.b.ii
Padé Filtering of a deterministic oscillatory signal of length $N = 20$. Density of the poles on $\mathbb{C}$ of a set of $\mathcal{M} = 5000$ draws of $[10/10]$ PA's. There are two peaks located at the resonances. Fig (ii) gives a landscape view of these two peaks.

Figures 14.c.i and 14.c.ii
Padé Filtering of a random white gaussian process of length $N = 20$. Density of the poles of a set of $\mathcal{M} = 5000$ draws of $[10/10]$ PA's. Fig. (ii): landscape view.

Figures 14.d.i and 14.d.ii
Padé Filtering of a random AR(1) gaussian process of length $N = 20$. Density of the poles of a set of $\mathcal{M} = 5000$ draws of $[10/10]$ PA's. The two white dots indicate the resonance and the inverse resonance. They 'attract' part of the poles, both on the real axis and on the circle. Fig. (ii): landscape view.

Figures 14.d.iii and 14.d.iv
Padé Filtering of a random AR(2) gaussian process of length $N = 20$. Density of the poles of a set of $\mathcal{M} = 5000$ draws of $[10/10]$ PA's. The four white dots are at the resonances and the inverse resonances. They operate a confinement of a fraction of the poles situated on the circle. Fig. (iv): landscape view.

Figures 14.e    Padé Filtering of data of length $N = 20$ which are the superposition of a deterministic cosine signal and a white gaussian noise. Density of the poles of a set of $\mathcal{M} = 5000$ draws of $[10/10]$ PA's. For Fig. 14.e.(i) the signal and the noise have the same amplitude, for fig. 14.e.(iii) the signal is twice larger and Fig. 14.e.(v) it is 10 times larger. In all cases the circular invariance of the pure white random case is broken. Fig. 14.e.(ii), (iv), (vi) give the perspective view for respectively Fig. 14. e.(i), (iii), (v).

Figure 15.1    GW chirp signal in the late inspiral phase of coalescing binaries. The local amplitude and frequency increase to potentially become infinite.

Figure 15.2    Padé Filtering of the chirp signal depicted in Figure 15.1. PA $[49/50]$. Position of the poles.

Figure 15.3    As Figure 15.2. Position of the zeros.

Figure 15.4    GW decreasing oscillatory signal in the ringdown phase after the coalescence of two bodies. The local amplitude exponentially decreases while the frequency Is kept constant.

Figure 15.5    Padé Filtering of the ringdown signal depicted in Fig. 15.4. Numerical illustration of the theoretical findings. The $[49/50]$ PA reduces to a $[1/2]$ fraction. The zero is in green, the two poles are in red; with a zoom on their position.

Figure 15.6    Padé Filtering (poles (i) and zeros (ii)) of a model mimicking a pulsar continuous GW signal. PA $[50/50]$, frequency $F = 20\,Hz$, sampling frequency $F_S = 200\,Hz$. On the unit circle the positions of the poles and zeros are closed to each other, while the peaks at the origin are extremely closed and cancel each other.

Figure 16.1.i    The vertical component of the micro-seismicity Virgo is subjected to is plotted against time for five consecutive working days. The gross 1-day periodicity points the role of anthropic disturbances.

Figure 16.1.ii    The vertical component of the motion of the 0[th] stage (resp. 7[th] stage) of the suspension of a Virgo mirror is plotted in green (resp. blue) against time for five consecutive working days. The two signals have a very different order of magnitude but their profiles are related as revealed by Fourier analysis. (Fig. 16.2.(i))

Figure 16.2.i    Energy spectrum of the green (resp. blue) mirror suspension motion signals of Fig. 16.1.(ii) in common arbitrary unit. The two plots exhibit the same bumps and a peak at very low frequency.

Figure 16.2.ii    Energy spectrum of the micro-seismicity signal plotted in Fig. 16.1.(i). It exhibits a peak at very low frequency.

Figure 16.3.i    In blue (resp. red) the cross-correlation function of the seismicity and the 0[th] stage (resp. 7[th] stage) of a Virgo mirror suspension. In green the cross-correlation function of the 0[th] stage and the 7[th] stage.

Figure 16.3.ii    In red the auto-correlation function of the seismicity. In blue (resp. green) the auto-correlation function of the 0[th] (resp. 7[th]) stage motion.

Figure 16.4.i    Padé Filtering of the seismicity channel: distribution of the poles (resp zeros) in red (resp. green). PA $[20/20]$.

Figure 16.4.ii    Padé Filtering of the suspension's 0[th] stage motion: distribution of the poles (resp. zeros) in red (resp. green). PA $[20/20]$.

Figure 16.5    Skeleton of the analytic structure (five poles in red, five zeros in green) of the z-transform of the operator relating seismicity and 0[th] stage of the suspension. This skeleton is obtained in erasing all the doublets in the Padé Filtering presented in Fig. 16.4.(i,ii).